\documentclass[fleqn,usenatbib]{mnras}

\usepackage{amssymb}

\usepackage[T1]{fontenc}

\DeclareRobustCommand{\VAN}[3]{#2}
\let\VANthebibliography\thebibliography
\def\thebibliography{\DeclareRobustCommand{\VAN}[3]{##3}\VANthebibliography}

\usepackage{graphicx}	
\usepackage{amsmath}	
\usepackage{amssymb}	

\usepackage{newtxtext,newtxmath}

\title[CNN identification of post-mergers in UNIONS using IllustrisTNG]{Convolutional neural network identification of galaxy post-mergers in UNIONS using IllustrisTNG}

\author[R. W. Bickley et al.]{Robert W. Bickley,$^{1}$\thanks{E-mail: rbickley@uvic.ca}
Connor Bottrell,$^{2,1}$
Maan H. Hani,$^{3,1,\thanks{Vanier Scholar}}$
Sara L. Ellison,$^{1}$
\newauthor
Hossen Teimoorinia,$^{4}$
Kwang Moo Yi,$^{5,1}$
Scott Wilkinson,$^{1}$
Stephen Gwyn,$^{6}$
\newauthor
Michael J. Hudson$^{7,8,9}$
\\
$^{1}$Department of Physics and Astronomy, University of Victoria, Victoria, British Columbia V8P 1A1, Canada\\
$^{2}$Kavli IPMU (WPI), UTIAS, The University of Tokyo, Kashiwa, Chiba 277-8583, Japan\\
$^{3}$Department of Physics and Astronomy, McMaster University, Hamilton Ontario L8S 4M1, Canada\\
$^{4}$National Research Council of Canada, 5071 West Saanich Road, Victoria, British Columbia V9E 2E7, Canada\\
$^{5}$Department of Computer Science, University of British Columbia, 2366 Main Mall \#201, Vancouver, British Columbia V6T 1Z4, Canada\\
$^{6}$Canadian Astronomy Data Centre, NRC Herzberg, 5071 West Saanich Road, Victoria, BC, V9E 2E7, Canada\\
$^{7}$Department of Physics and Astronomy, University of Waterloo, 200 University Ave W, Waterloo, ON N2L 3G1, Canada\\
$^{8}$Waterloo Centre for Astrophysics, University of Waterloo, 200 University Ave W, Waterloo, ON N2L 3G1, Canada\\
$^{9}$Perimeter Institute for Theoretical Physics, 31 Caroline St. North, Waterloo, ON N2L 2Y5, Canada
}

\date{Accepted XXX. Received YYY; in original form ZZZ}

\pubyear{2021}

\begin{document}
\label{firstpage}
\pagerange{\pageref{firstpage}--\pageref{lastpage}}
\maketitle

\begin{abstract}
The Canada-France Imaging Survey (CFIS) will consist of deep, high-resolution r-band imaging over \textasciitilde5000 square degrees of the sky, representing a first-rate opportunity to identify recently-merged galaxies. Due to the large number of galaxies in CFIS, we investigate the use of a convolutional neural network (CNN) for automated merger classification. Training samples of post-merger and isolated galaxy images are generated from the IllustrisTNG simulation processed with the observational realism code \textsc{RealSim}. The CNN's overall classification accuracy is 88 percent, remaining stable over a wide range of intrinsic and environmental parameters. We generate a mock galaxy survey from IllustrisTNG in order to explore the expected purity of post-merger samples identified by the CNN. Despite the CNN's good performance in training, the intrinsic rarity of post-mergers leads to a sample that is only \textasciitilde6 percent pure when the default decision threshold is used. We investigate trade-offs in purity and completeness with a variable decision threshold and find that we recover the statistical distribution of merger-induced star formation rate enhancements. Finally, the performance of the CNN is compared with both traditional automated methods and human classifiers. The CNN is shown to outperform Gini-M20 and asymmetry methods by an order of magnitude in post-merger sample purity on the mock survey data. Although the CNN outperforms the human classifiers on sample completeness, the purity of the post-merger sample identified by humans is frequently higher, indicating that a hybrid approach to classifications may be an effective solution to merger classifications in large surveys.
\end{abstract}




\section{Introduction}


Observations and theory alike suggest that mergers, as the principle nodes of assembly, alter the evolutionary trajectory of galaxies (\citealp{1978MNRAS.183..341W}; \citealp{1993MNRAS.262..627L}). Further, the ensemble of ways in which a merger recasts a galaxy is unique among transformative processes; mergers are capable of inducing simultaneous changes in the visual and underlying properties of the host on relatively short timescales (e.g. \citealp{2008MNRAS.383...93B}; \citealp{2008ApJ...675.1095J}). N-body simulations (e.g. \citealp{1972ApJ...178..623T}; \citealp{Conselice_2006}) and hydrodynamical simulations of galaxy mergers (e.g. \citealp{2008MNRAS.391.1137L}) produce many of the morphological signatures, including stellar shells, bridges, and tidal tails, that are seen in observed mergers (\citealp{2010MNRAS.401.1043D}; \citealp{2015ApJS..221...11K}; \citealp{2017MNRAS.464.4420S}).

Theoretical models also predict a number of non-morphological merger-induced effects (e.g. \citealp{1972ApJ...178..623T}; \citealp{2005MNRAS.361..776S}; \citealp{2008AN....329..952D}; \citealp{2019MNRAS.485.1320M}; \citealp{2020MNRAS.494.4969P}; \citealp{2020MNRAS.493.3716H}), many of which are supported by observations. These include central starbursts associated with an influx of cold gas (\citealp{2008AJ....135.1877E, 2013MNRAS.435.3627E}; \citealp{2004MNRAS.355..874N}; \citealp{2014MNRAS.437.2137S}; \citealp{2015HiA....16..326K}; \citealp{2019MNRAS.482L..55T}), which may additionally lead to enhanced accretion onto the central super-massive black hole and triggering of an active galactic nucleus (AGN; e.g., \citealp{2011gafo.confE..83K}, \citealp{2014MNRAS.441.1297S}, \citealp{2011MNRAS.418.2043E, 2019MNRAS.487.2491E}). Surges in central star formation as well as the AGN may in turn contribute to outflows from the galaxy, enriching the circum-galactic medium (\citealp{2015MNRAS.449.3263J}; \citealp{2018MNRAS.475.1160H}) and extending the impact of a merger beyond the scale of the participant galaxies. Mergers also stand to transform the kinematics of galaxies (\citealp{1967MNRAS.136..101L}; \citealp{1977egsp.conf..401T}; \citealp{1983MNRAS.205.1009N}; \citealp{1992ApJ...400..460H}; \citealp{2003ApJ...597..893N}; \citealp{2006ApJ...645..986R}; \citealp{2009MNRAS.397.1202J}; \citealp{2014MNRAS.440L..66B}; \citealp{2018MNRAS.478.3994C}; \citealp{2018MNRAS.475.1160H}), and after an initial starburst, may serve to accelerate a galaxy's star formation schedule towards a more permanent state of quiescence (\citealp{1988ApJ...325...74S}; \citealp{2006ApJS..163....1H}; \citealp{2014ApJ...792...84Y}; \citealp{2021MNRAS.XXX.XXXXQ}).

Observational tests of these theoretical predictions are predicated on the initial identification of a merger sample. In a binary galaxy merger, the pair stage immediately precedes galaxy coalescence. The pair phase is relatively straightforward to identify visually (e.g. \citealp{2007ApJS..172..329K}; \citealp{2005ApJ...625..621B}; \citealp{1998ApJ...499..112B}; \citealp{2010MNRAS.401.1043D}), or by searching for galaxies close in angular separation and line-of-sight radial velocity. Numerous studies of close pairs identified in large spectroscopic surveys have demonstrated that automated spectroscopic identification is an effective method to study the impact of the pre-coalescence interaction in large statistical samples of galaxies (e.g. \citealp{2000ApJ...536..153P}; \citealp{2000ApJ...530..660B}; \citealp{2004ApJ...617L...9L}; \citealp{2005AJ....130.1516D}; \citealp{2008ASPC..399..298L}).

The study of post-mergers, that is, merger remnants that have combined into new, sovereign entities, complements investigations of the pair phase by extending observations into the post-coalescence era of the interaction. Because spectra can no longer be used to identify post-merger galaxies, post-merger samples have been limited in size due to the time commitment associated with robust visual post-merger identification by human classifiers (e.g. \textasciitilde100 post-mergers in \citealp{2013MNRAS.435.3627E}). Nevertheless, such studies have allowed for photometric and spectroscopic measurement of merger-induced evolutionary effects and detailed kinematic and profile studies via cutting-edge integral field spectroscopy (IFS) instruments (e.g. \citealp{2019MNRAS.482L..55T}; \citealp{2015A&A...582A..21B}; \citealp{2019ApJ...881..119P}, \citealp{2020arXiv200902974W}). Moreover, \citet{2013MNRAS.435.3627E} demonstrated that the post-coalescence era is the most transformative stage of the merger sequence, with the strongest changes in SFR, metallicity and AGN fraction. However, the identification of galaxies that have fully undergone a recent merger presents a greater initial challenge for several reasons. First, the visual signatures of a post-coalescence galaxy are relatively short-lived, and can be difficult to extricate from other nominal irregularities in morphology (e.g. \citealp{2008MNRAS.391.1137L,2011ApJ...742..103L}). Even when merger-induced morphological phenomena are present, they are often faint, and can be easily be overlooked on account of survey artifacts, bright foreground objects, and observational noise. Further, unlike the identification of pairs, the identification of post-mergers is a more challenging prospect for automation. Irregular morphological features among isolated galaxies have the potential to delude human and automated classifiers alike into assigning false post-merger labels. Therefore, robust identification of post-merger galaxies requires the fulfillment of two broad criteria: (1) images that are sufficiently well-resolved and deep (e.g. \citealp{2019MNRAS.486..390B}) to reveal the features typically associated with mergers, and (2) an identification method that is insensitive to features that are irrelevant to a galaxy's status as either a post-merger, or a non-post-merger.

The Canada-France Imaging Survey (CFIS) readily fulfills the first of the aforementioned criteria of imaging depth, width and seeing quality. Even with conservative estimates of post-merger occurrence rates in the low-redshift Universe, the survey, will include r-band imaging with ~0.6 arcsecond seeing for over 5000 square degrees of the sky, is likely to contain thousands of post-mergers (\citealp{1993MNRAS.262..627L}; \citealp{2011ApJ...742..103L}; \citealp{2012ApJ...747...34B}; \citealp{2014MNRAS.445.1157C}; \citealp{2015MNRAS.449...49R}; \citealp{2018MNRAS.480.2266M}). Moreover, the survey's 5-$\sigma$ point-source depth (24.85 mag in the r band for the MegaCam wide-field optical imager) is sufficient to capture the low-surface brightness features associated with mergers in exquisite detail. Better still, \textasciitilde3300 square degrees of the survey overlap with the SDSS Baryon Oscillation Spectroscopic Survey (BOSS) (\citealp{2013AJ....145...10D}) footprint, paving the way for subsequent cross-survey spectroscopic characterization of a well-defined post-merger population identified by CFIS.

The volume of galaxies imaged by CFIS, which contains the promise of a statistically large sample of post-mergers, presents a challenge of its own. Expert classifications by an individual trained to disentangle the visual characteristics of post-mergers from noise, obstruction, and nominal morphological deviation, are widely considered to be of formidable quality (e.g. \citealp{2010ApJS..186..427N}, \citealp{2013MNRAS.435.3627E}). However, classifications of >100,000 CFIS galaxies by an individual, or even a group of individuals, would represent months of uninterrupted effort. Visual classifications are also difficult to reproduce reliably. By contrast, automated statistical methods of classification (\citealp{Conselice_2003}; \citealp{2004AJ....128..163L}; \citealp{2016MNRAS.456.3032P}; \citealp{2019MNRAS.483.4140R}; \citealp{2019ApJ...872...76N}), many of which are motivated by physics, can be calculated nearly instantaneously for a single image, and return morphological parameters including asymmetry, Gini, and M20, which may provide clues, albeit ones that are highly sensitive to survey irregularities, resolution, and surface brightness limits (e.g. \citealp{2014A&A...566A..97J,2019MNRAS.486..390B}), as to the morphological status of a large population of galaxies.

Deep learning classification models, particularly convolutional neural networks (CNNs), have gained favour in the last decade for image classification problems of all kinds, both within and outside of astronomy (\citealp{2015ApJS..221....8H}; \citealp{2018MNRAS.476.3661D}; \citealp{2019ApJS..243...17J}; \citealp{2019MNRAS.484...93D}; \citealp{2019ApJ...876...82N}; \citealp{2019ascl.soft06012H}; \citealp{2019MNRAS.489.1859H}), and offer a path forward for the second aforementioned criterion. CNNs are capable of identifying features, and feature combinations, with a great deal of nuance, and have shown promise as a means of enhancing the quality of morphological labels assigned to galaxies. Consequently, CNNs have already been identified and exploited as a means to effectively harvest mergers from a range of imaging surveys (e.g. \citealp{2018MNRAS.479..415A}; \citealp{2019MNRAS.483.2968W}, \citealp{2019A&A...626A..49P}, \citealp{2020ApJ...895..115F}, \citealp{2020arXiv200902974W}).

However, neural networks lack an understanding of physics, and so the morphological classifications that fall out of a CNN are only trustworthy insofar as the relevant features in the training data are representative of the problem at hand. For post-merger identification in particular, this means that realistic and diverse examples of post-mergers must be offered to a CNN so that it can begin to make informed generalizations about the post-merger and non-post-merger classes, and infer merger status from images. Additionally, the CNN must be prepared to encounter a range of features that are explicitly irrelevant to the morphological status of a galaxy: survey-related artifacts and noise; the presence or lack of bright foreground objects; the presence of neighbour galaxies; or even the presence of companion galaxies when the target is in the pair phase, which ought not to be included in a sample of fully-coalesced galaxies. Previous work by \citet{2019MNRAS.490.5390B} and by \citet{2019MNRAS.489.1859H} have shown that observational realism, that is, observationally accurate levels of noise, and the presence of observational artifacts in representative quantities, is of paramount importance for the success of a CNN in the context of morphological classification.

In addition to a realistic and equitable distribution of observational effects in the training set of a CNN, it is also crucial to prepare the network for the diversity of situations in which a post-merger might be found. Due to the ubiquity of galaxy mergers within the framework of hierarchical galaxy assembly and evolution, galaxies of all types, regardless of their initial morphology, dynamical state, size, or neighbourhood density, might undergo mergers. We seek to account for this fact in training as well, by preparing post-merger and non-post-merger samples that are matched in redshift and environmental parameters, and which encompass the diversity of morphologies and evolutionary histories of real galaxies. This prods the network towards the conclusion that these parameters are unlikely to be relevant to the classification task at hand.

A further factor affecting the performance of a CNN is the quality of the image labels provided during training. The source of these labels must be well-considered, as their trustworthiness places an inflexible upper limit on the network's ability to learn about the problem, and to make conclusions that are rooted in the physics that give rise to the morphologies visible in galaxy images. Human-provided labels, therefore, are not ideal; regardless of volume or quantity of experience in image classification, biases and strategies employed by human classifiers are frequently detrimental to the purity and overall quality of a training sample. Cosmological simulations, which host manifold populations of galaxies, circumvent the issue of visual biases by virtue of the fact that the ground truth - whether or not a given galaxy has recently coalesced after a merger - is unambiguously accessible. Even deploying a very strict definition of post-merger status, the 100-1 run of the IllustrisTNG cosmological magnetohydrodynamical simulations (\citealp{2018MNRAS.480.5113M}; \citealp{2018MNRAS.477.1206N}; \citealp{2018MNRAS.475..624N}; \citealp{2018MNRAS.475..648P}; \citealp{2018MNRAS.475..676S}; \citealp{2019ComAC...6....2N}) includes more than 2,000 post-merger galaxies in a range of environments, derived from a highly heterogeneous group of progenitors, and representative of galaxies in the observed Universe. \citet{2019MNRAS.483.4140R} reported good agreement (\textasciitilde1-$\sigma$ for all parameters studied) between the optical morphologies of IllustrisTNG galaxies processed with \textsc{skirt} radiative transfer\footnote{skirt.ugent.be} (\citealp{2011ascl.soft09003B}; \citealp{2015A&C.....9...20C}), and real galaxies observed by Pan-STARRS (\citealp{2016arXiv161205560C}). To the extent that the morphological and evolutionary characteristics of the galaxies spawned by IllustrisTNG are faithful to those found in the Universe, a CNN trained on sufficiently realistic images derived from these simulated galaxies will be prepared to identify post-mergers as seen by CFIS.

In this paper, we: (1) construct a training sample of statistically representative post-merger and non-post-merger images of synthetic galaxies from the IllustrisTNG simulation, processed with statistical and observational realism faithful to CFIS (Section~\ref{Methods}), (2) train a CNN on the image sample so constructed, and evaluate its performance by traditional machine learning metrics as well as novel methods to assess its consistency as a function of the constitutive and environmental characteristics of the target galaxies (Sections~\ref{Test Set Properties} and \ref{Performance Trends}), (3) use the CNN as a source of post-merger labels within the context of a large mock survey (Section~\ref{Mock Survey}), and (4) compare the performance of the CNN to that of traditional automated methods, as well as a group of volunteer visual classifiers, in order to identify the most efficient empirical method that balances automation and post-merger sample purity, to be applied to CFIS in a forthcoming paper (Sections~\ref{Comparison to Automated Methods} and \ref{Visual Classification}).

\section{Methods}
\label{Methods}

In the following section, we describe the IllustrisTNG simulations and the selection criteria for the post-merger and control classes, (Sections~\ref{IllustrisTNG 100-1}, \ref{Post-mergers}, and \ref{Controls-Environment}, respectively), the synthetic observation pipeline applied to the galaxies in order to generate the image data (Section~\ref{Synthetic Observations}), and the architecture and hyperparameters deployed in the CNN that will be trained on the images (Section~\ref{CNN Architecture}).

\subsection{Simulations}
\label{Simulations}

\subsubsection{IllustrisTNG 100-1}
\label{IllustrisTNG 100-1}

In order to acquire galaxy samples that are representative of the observable Universe, we turn to large-box cosmological magnetohydrodynamical simulations. Specifically, we utilise the 100-1 run of the IllustrisTNG simulations (\citealp{2018MNRAS.480.5113M}; \citealp{2018MNRAS.477.1206N}; \citealp{2018MNRAS.475..624N}; \citealp{2018MNRAS.475..648P}; \citealp{2018MNRAS.475..676S}; \citealp{2019ComAC...6....2N}), the volume of which is cubic in shape with a side length of 110.7 Mpc. In 100-1, the highest-resolution run of this simulation volume, the baryonic matter resolution is $\mathrm{1.4x10^6}$ \(\textup{M}_\odot\), such that a galaxy meeting the minimum stellar mass criterion for this work ($\mathrm{10^{10}}$ \(\textup{M}_\odot\)) would be resolved with \textasciitilde$10^{4}$ star particles. We select galaxies for training and evaluation from simulation snapshots 50-99, or $z$ = 1 through $z$ = 0; each snapshot corresponds to \textasciitilde160 Myr.

\subsubsection{Post-mergers}
\label{Post-mergers}

Post-mergers are identified using the merger trees created by \textsc{Sublink} (\citealp{2015MNRAS.449...49R}), following the methodology of \citet{2020MNRAS.493.3716H}. We consider only mergers with stellar mass ratios $\mu \geq 0.1$, that occurred at $z \leq 1$ (snapshot 50 of the simulation), and for which the merger remnant is in the stellar mass range $\mathrm{10^{10}-10^{12}}$ \(\textup{M}_\odot\). In each simulation snapshot, each galaxy is given a post-merger time, $\mathrm{T}_\mathrm{Postmerger}$, denoting the time since the most recent merger along its tree. We then define our post-merger sample as those galaxies whose $\mathrm{T}_\mathrm{Postmerger}$ = 0 (i.e., galaxies for which a merger elapsed between the current and previous simulation snapshots). These selection criteria are designed to maximize the volume and diversity of well-resolved post-merger properties, and yield a sample of 2332 post-merger galaxies (see \citealp{2020MNRAS.493.3716H} for more details).

\subsubsection{Controls}
\label{Controls-Environment}

We seek to prepare a post-merger identification tool that is capable of distinguishing observed post-mergers, encompassing a range of diverse characteristics, from similarly varied observations of non-post-merger galaxies. We therefore structure our work as a binary classification problem, in which our CNN will identify a given galaxy as either a post-merger, or a non-post-merger. In order to be able to label non-post-mergers correctly, the CNN must be trained on a representative sample of non-post-merger galaxies. Our control group identification methodology aims to construct such a sample.

In order to generate a control group of non post-mergers, we adapt the approach of \citet{2016MNRAS.461.2589P} using IllustrisTNG metadata. Four unique control galaxies are identified for each post-merger galaxy via a growing parameter search matching in stellar mass, simulation snapshot number (i.e. simulation redshift), the distance in kpc to the nearest and second-nearest neighbour galaxies with $\mathrm{M} \geq 0.1*\mathrm{M}_\mathrm{host}$ ($\mathrm{r_{1}}$ and $\mathrm{r_{2}}$, respectively), and $\mathrm{N_{2}}$, the number of galaxies within 2 Mpc. Additionally, the same total stellar mass cuts on the post-merger sample are applied to control pool. Control galaxies must not have undergone a merger in the last 2 Gyr (see also \citealp{2020MNRAS.493.3716H}). We identify the 2 Gyr lower bound for the control group as a conservative choice, as the morphological signatures of a merger are only expected to last $\lesssim$ 1 Gyr (e.g. \citealp{Conselice_2006}; \citealp{2008MNRAS.391.1137L}). The continued observability of merger features may be further limited by imaging depth or simulation resolution. Additionally, softening the threshold to 1.5 Gyr does not does not qualitatively change our results.

The default matching tolerances are 0.1 dex in stellar mass, 1 simulation snapshot prior to or later than that of the post-merger, and $\pm10\%$ for $\mathrm{r_{1}}$, $\mathrm{r_{2}}$, and $\mathrm{N_{2}}$. A snapshot (i.e. simulation redshift) tolerance is allowed, as each snapshot for ${1\leq z \leq 0}$ only corresponds to a ${\Delta z}$ of $\leq 0.05$, and morphological evolution is therefore limited. Even these relatively narrow parameters usually yield multiple eligible controls, from which one is selected at random. Once a galaxy is selected for use as a control, it is marked as such, preventing it from being re-selected as a control for a different post-merger. Disallowing replacement in the training sample ensures that each selected control galaxy is unique. In the case where no eligible controls are found, the tolerances are gradually increased; the snapshot number tolerance is increased by one, the stellar mass tolerance in dex is increased by a factor of 1.5, and the $\mathrm{r_{1}}$, $\mathrm{r_{2}}$, and $\mathrm{N_{2}}$ tolerances are also increased by a factor of 1.5. This is repeated as necessary in four serialized rounds, such that one control is found for each post-merger before repeating the procedure to yield 4 unique controls for each post-merger. This ensures that the pool of potential control galaxies is not depleted by the post-mergers that find their controls first. In the full control sample, \textasciitilde44\% are found using the default match tolerances, \textasciitilde24\% required the tolerances to be softened exactly once, \textasciitilde19\% required two growths, \textasciitilde10\% required three growths, and < 4\% required more than three growths in parameter space.

\begin{figure*}
\includegraphics[width=\textwidth]{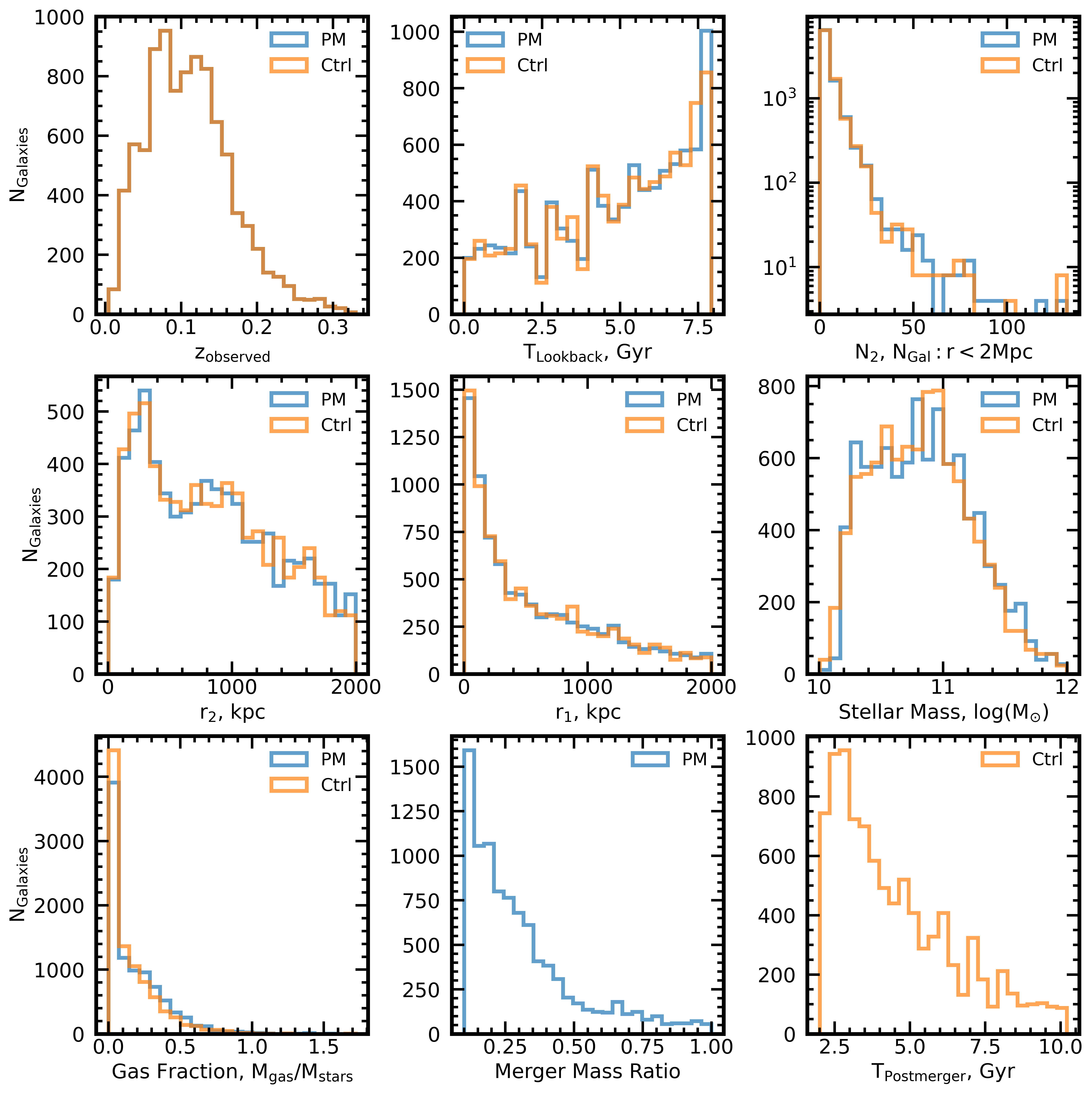}
\caption{The constitutive, environmental, and observational statistics of the post-merger and control samples. Redshift is applied in mirrored fashion on-the-fly for a post-merger, control pair, and so their distributions in the top left panel match exactly. The two samples track each other in lookback time, $\mathrm{N_{2}}$, $\mathrm{r_{1}}$, $\mathrm{r_{2}}$, and stellar mass as a result of our matching procedure. The similarity of the populations in gas fraction is a consequence of the parameters that are explicitly matched. Mass ratio and time-since-merger statistics are also shown for the relevant groups.}
\label{fig:multi_hist}
\end{figure*}

Figure~\ref{fig:multi_hist} shows the distribution of properties of the post-merger and control samples (only the first round of matched controls is shown so that the histograms contain equal numbers of post-mergers and controls). The mock observed redshift distributions (top left panel) match exactly as a consequence of the mirrored treatment of mock observed redshift, detailed in Section~\ref{RealSimCFIS}, whereas simulation lookback time (top centre panel) is a matched parameter from IllustrisTNG. Figure~\ref{fig:multi_hist} shows that the matching process has done an excellent job of mirroring the distributions of stellar mass, $\mathrm{r_{1}}$, $\mathrm{r_{2}}$ and $\mathrm{N_{2}}$ between the post-mergers and controls. The similarity of the post-merger and control samples in gas fraction (bottom left panel), calculated as the ratio of gas mass to stellar mass within twice the galaxy's half-mass radius, is a secondary effect of the parameters on which the two are explicitly matched: specifically, gas fraction exhibits a strong negative correlation with stellar mass. Figure~\ref{fig:multi_hist} also shows the distribution of the stellar mass ratio of the post-mergers' progenitors, and $\mathrm{T}_\mathrm{Postmerger}$ of the controls, where the latter is zero by definition for the post-mergers and > 2 Gyr for the controls. 85\% of the post-mergers have a mass ratio of $\mu$ < 0.5, and many galaxies in the control sample have undergone mergers of their own within the past 5 Gyr, although the minimum $\mathrm{T}_\mathrm{Postmerger}$ cutoff for control galaxies ensures that no visual signatures of a merger remain.

\subsection{Synthetic observations}
\label{Synthetic Observations}

\citet{2019MNRAS.490.5390B} studied the importance of observational realism to the reliability of CNN predictions of galaxy merger stage, and found that mismatched levels of realism in training and test images would incur a significant penalty in performance. Further, training data with appropriate observational realism was found to be more important to CNN performance than the treatment of either colour or radiative transfer. Because we aim to construct a post-merger identification tool for the specific case of CFIS, we combine the survey's available image data and metadata in order to construct a sample of synthetic training images that patterns itself after the statistical properties and observational qualities of the survey. In addition, we forgo the use of radiative transfer in favour of unprocessed stellar maps as a starting point for our mock observations in order to limit computational expense as proposed in \citet{2019MNRAS.490.5390B}.

\subsubsection{UNIONS (CFIS) Observations}
\label{UNIONS (CFIS) Observations}

A pre-requisite for any survey of mergers is high quality imaging over a large sky area. The Ultraviolet Near Infrared Optical Northern Survey (UNIONS) collaboration is a new consortium of wide field imaging surveys of the northern hemisphere and represents an excellent opportunity for merger searches. UNIONS consists of CFIS conducted at the 3.6-meter CFHT on Maunakea, members of the Pan-STARRS team, and the Wide Imaging with Subaru HyperSuprimeCam of the Euclid Sky (WISHES) team. CFHT/CFIS is obtaining deep u and r bands; PanSTARRS is obtaining deep i and moderate-deep z band imaging, and Subaru/WISHES is obtaining deep z band imaging. These independent efforts are directed, in part, to securing optical imaging to complement the Euclid space mission, although UNIONS is a separate consortium aimed at maximizing the science return of these large and deep surveys of the northern skies. In the construction of our mock observation images for this contribution, we make use of the CFIS r-band data only.

The observing pattern employed by CFIS uses three single-exposure visits with FOV offsets in between for optimal astrometric and photometric calibration with respect to observing conditions. This also ensures that the entire survey footprint will be visited for at least two exposures. After raw images are collected by CFHT, they are detrended (i.e. the bias is removed and the images are flat-fielded using night sky flats) with the software package MegaPipe (\citealp{2008PASP..120..212G}). The images are next astrometrically calibrated using Gaia DR2 (\citealp{2016A&A...595A...1G,2018A&A...616A...1G}) as a reference frame. Pan-STARRS 3pi r-band photometry (\citealp{2016AAS...22732407C}) is used to generate a run-by-run differential calibration across the MegaCam mosaic, and an image-by-image absolute calibration. Finally, the individual images are stacked onto an evenly spaced grid of 0.5-degree-square tiles.

\subsubsection{Stellar maps}
\label{Stellar Maps}

Synthetic observations for this work begin with stellar mass maps rendered from the simulation for each selected galaxy. Each galaxy is observed by four different camera angles, located at the vertices of a tetrahedron which is aligned with the simulation box, with the galaxy at its geometric centre. As a result, the first camera angle looks directly down on each galaxy perpendicular to the top of the simulation box, and the remaining three are inclined upwards from horizontal. These camera angles are chosen to capture distinct projected morphologies for each galaxy so that the CNN will be able to study a greater number of unique examples. Each map is 100 kpc and 2048 pixels on a side. These stellar mass maps are then normalized using the galaxy's r-band absolute magnitude from the \citet{2019ComAC...6....2N} stellar photometrics tables to produce a pre-cosmology image of intrinsic surface brightness. Single-band photometry ensures that the features learned by the CNN are morphological, and not biased by higher-order information such as colour or starburst identification. The top row of Figures~\ref{fig:pm_mosaic} and \ref{fig:ctrl_mosaic} show selected stellar mass maps for post-mergers and their controls, respectively.

\subsubsection{\textsc{RealSim-CFIS}}
\label{RealSimCFIS}

The pristine surface brightness images are next convolved with \textsc{RealSim-CFIS}\footnote{github.com/cbottrell/RealSimCFIS}, a custom version of \textsc{RealSim}\footnote{github.com/cbottrell/RealSim}, originally detailed in \citet{2019MNRAS.490.5390B}. The code was originally developed to construct synthetic images of simulated phenomena as they might be observed by SDSS. In both versions of \textsc{RealSim}, the noise, resolution, and sky insertion positions for each mock observation are selected to match the statistics of sky brightness, seeing, image artifacts, and projected environment (i.e. crowding) for a catalog of real galaxies. Small modifications were required to adapt the code for synthetic CFIS observations: CFIS has an angular resolution of 0.187 arcsec/pixel, a factor of \textasciitilde2 higher than that of SDSS, allowing for greater preservation of the detail in the original unprocessed images. We also turn off the source Poisson noise feature from the original in the interest of simplicity. A given segment of the survey is imaged by a unique combination of potentially different charge-coupled devices (CCDs), and building a gain map when computing the Poisson statistics for a mock observation would require reverse-engineering of the gain for each pixel. \textsc{RealSim-CFIS} would also need to be modified to to accept such a gain map. We anticipate that any contribution by Poisson noise would be negligible compared to other simulated sources of noise applied later in the \textsc{RealSim-CFIS} pipeline. Before adding realistic observational effects to an image, it is first converted to SDSS-specific units of flux (nanomaggies) on a pixel-wise basis, with $10^{-0.4*(s-22.5)}$, where $s$ is a pixel's surface brightness value in magnitudes per square arcsecond.

\begin{figure*}
\includegraphics[width=\textwidth]{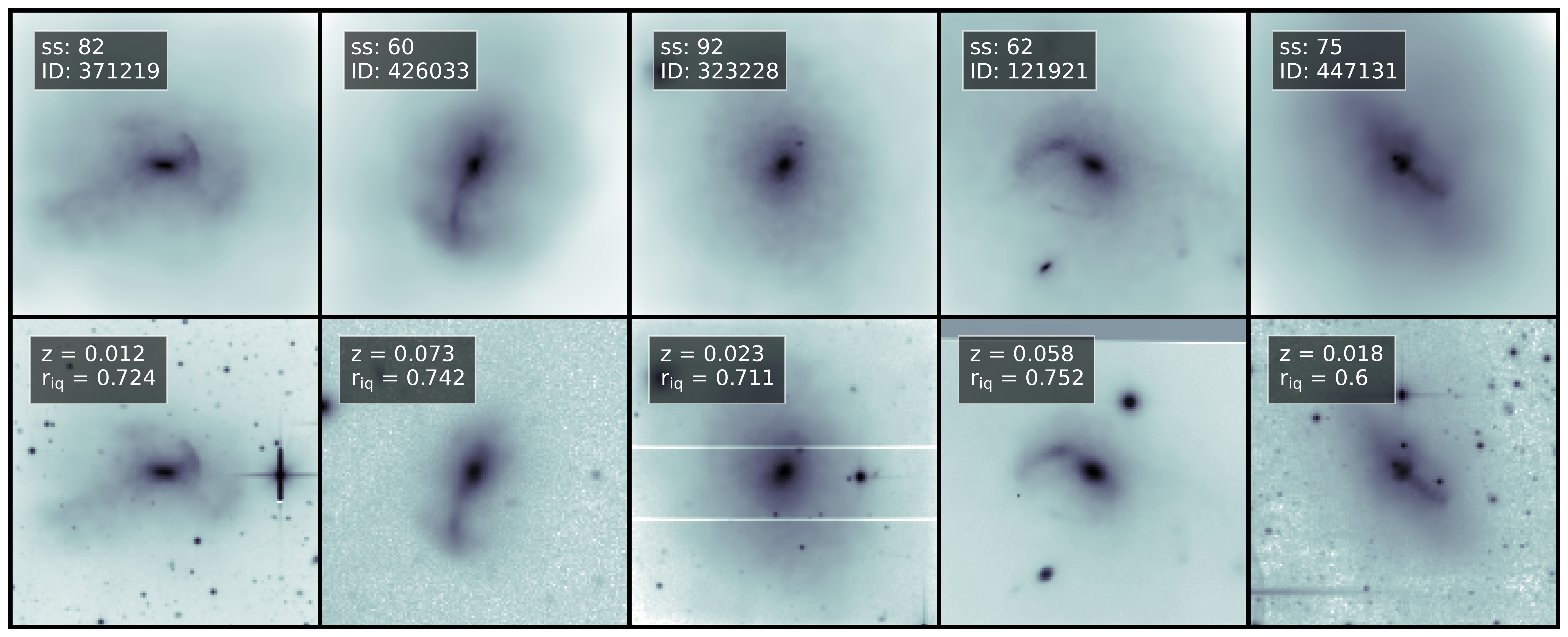}
\caption{Mosaic of 5 randomly selected post-mergers identified in the IllustrisTNG 100-1 run. Top row: stellar mass maps with the legend indicating the snapshot number and galaxy \textsc{Subfind} ID. Bottom row: Insertion into an actual CFIS r-band image with the legend indicating the insertion redshift and the image quality (in arcseconds) of the original image. Note how artifacts, such as saturated stars and CCD features are retained in the image.}
\label{fig:pm_mosaic}
\vspace*{\floatsep}
\includegraphics[width=\textwidth]{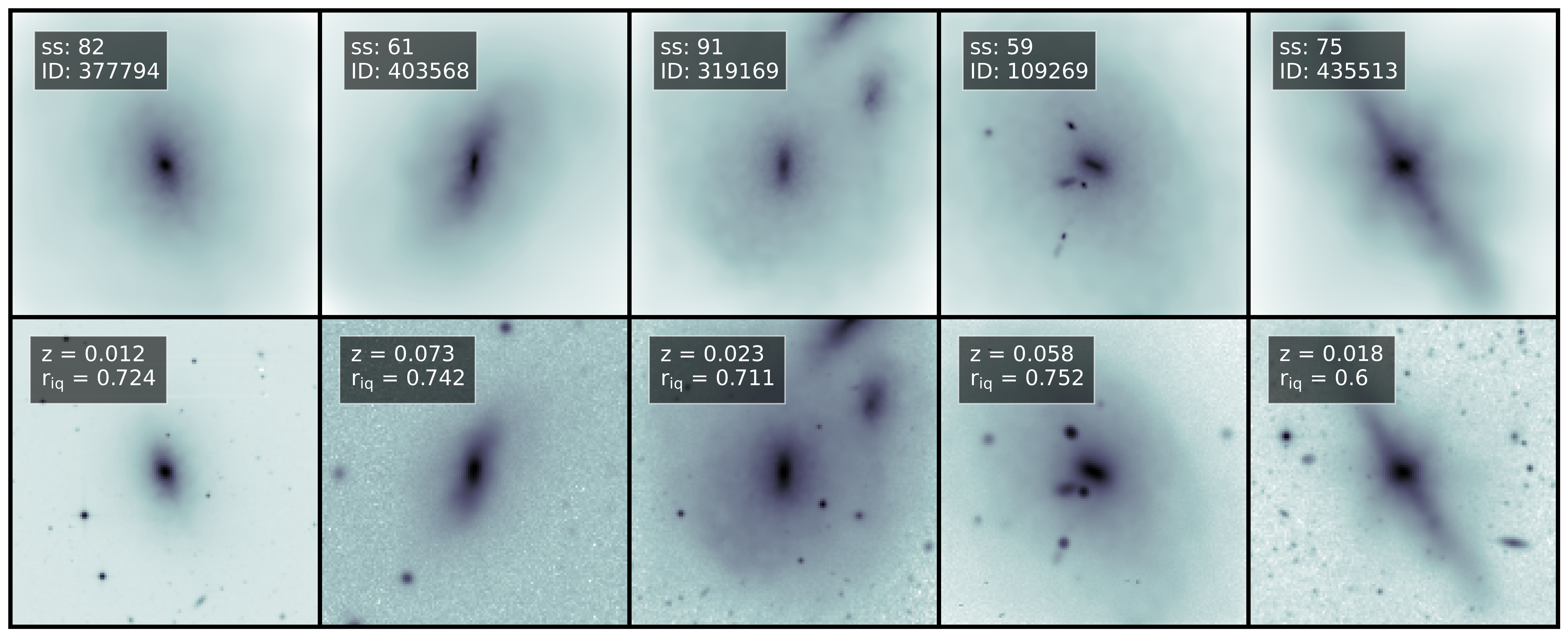}
\caption{As for Figure~\ref{fig:pm_mosaic} but for control galaxies corresponding to each of the five post-mergers shown in Figure~\ref{fig:pm_mosaic}. Note how the insertion redshifts are identical, as per our mock observation methodology. However, the seeing and image quality can differ between a given post-merger and control, as is usually the case in observational studies.}
\label{fig:ctrl_mosaic}
\end{figure*}

The layers of realism applied by the code are as follows:
\begin{itemize}
  \item \textit{Redshift Dimming}. Since future studies of the properties of post-mergers identified by our CNN in CFIS will make use of ancillary data from the SDSS, we choose insertion redshifts to match its observed redshift distribution. This redshift distribution is obtained by cross-matching CFIS Data Release 2 (DR2) object catalog with the SDSS Data Release 7 (DR7) with an angular separation tolerance of 2 arcseconds. Insertion redshifts are assigned to galaxies in the TNG post-merger sample by randomly drawing values from this catalog (with replacement). We cap the allowed mock observation redshift distribution at $z$ = 0.5 in order to rule out any galaxies whose ${z}_\mathrm{spec}$ may have been erroneously measured, or any high-redshift quasars that may have been included in the sample. The top left panel of Figure~\ref{fig:multi_hist} shows the mock observation redshift quantities obtained by drawing at random from the parent redshift distribution thus obtained. Each post-merger and associated control galaxy are assigned identical redshift selections. Post-merger, camera angle $(m,n)$ and control, camera-angle $(m,n)$ would therefore be mock observed at the same redshift. Once a redshift is selected, the image is realistically dimmed by a factor of $(1+z)^{-5}$. While cosmological surface brightness dimming accounts for a factor of $(1+z)^{-4}$ in bolometric surface brightness, an additional factor of $(1+z)^{-1}$ accounts for the dimming in a given bandpass.
  \smallskip
  \item \textit{Rebinning}. Once the angular size has been calculated from a given redshift and the physical size of the image (100 kpc), the redshift-dimmed image is rebinned to CFIS's actual CCD pixel scale in total flux-conserving fashion.
  \smallskip
  \item \textit{Point-Spread Function (PSF)}. The observational PSF for galaxies in CFIS is dominated by the effects of atmospheric seeing, and galaxies have associated PSF full-width at half-max (FWHM) values, recorded as $\mathrm{r}_\mathrm{iq}$, available as a metadata quantity in the CFIS catalogs. CFIS sky image data is separated into regularized 0.5 degree tiles, where each tile is the combination of several individual images. There is a small amount of variation in seeing across each tile resulting from the image mosaic. In order to model a spatially-variant PSF, we draw available PSF measurements on the CFIS tile where we intend to perform a mock observation. We next fit a gaussian function to the resulting distribution, and sample a value from the function at random. Consequently, we obtain a realistic and non-discrete approximation of the survey's atmospheric and instrumental seeing as a function of tile. This approach yields similar PSF distributions for the post-merger and control image sets because the underlying statistics are the same, but mirrored distributions are not enforced for post-mergers and controls, as a pair of galaxies matched on their physical properties and redshift are unlikely to be observed in identical conditions.
  \smallskip
  \item \textit{CFIS Images}. Finally, real CFIS skies are added to each image. Following the methods presented in \citet{2017MNRAS.467.1033B}, we employ a sky selection method that makes use of CFIS statistics in order to match the spatial distribution of real galaxies. In order to choose a CFIS tile, and in turn find a suitable survey location for a mock observation, we select a real proxy galaxy from the catalog of SDSS DR7 galaxies within the CFIS DR2 coverage. The proxy galaxy's tile will be used for the mock observation, and as a result, the CFIS tiles that are more densely populated with galaxies are more likely to be chosen for mock observations. In order to select a specific region of the tile for the mock observation, a large 11-arcminute-square cutout is generated, centered on the RA and DEC of the proxy galaxy. Next, Source Extractor (\citealp{1996A&AS..117..393B}) is used to identify an insertion location where the pixel corresponding to the centre of the output image cannot be flagged as a source. This allows for realistic overlap between the IllustrisTNG galaxy and survey objects. The chosen patch of sky is then added to the mock observation image, which is already at the CFIS CCD scale after rebinning. All of the original features and artifacts of the original CFIS image, such as saturated stars, missing sky coverage and CCD defects, are deliberately maintained in the final mock images through addition (in the case of high-flux artifacts) or masking (for zero-flux artifacts) in order to proportionally expose the CNN to phenomena that it is likely to find in a sample of CFIS images.
  
  In the rare case where the final image is dominated (>50\%) by zero-flux pixels due to lack of sky coverage, the synthetic observation is discarded, and a new observation of the same galaxy is attempted. The bottom row of panels in Figures~\ref{fig:pm_mosaic} and \ref{fig:ctrl_mosaic} show the same post-merger and control galaxies as in the upper row, but with all of the above described observational realism included.
  \smallskip
  \item \textit{Mock Observation Scheme}. Due to the relative scarcity of post-mergers in IllustrisTNG, we perform four unique mock observations of each of the four camera angles for every post-merger. Each post-merger mock observation (at each camera angle) consists of its own unique random draw from the redshift distribution and sky insertion location.  This results in 16 distinct images of each post-merger. Since there are four controls matched to each post-merger, each projected onto four camera angles, only one set of mock observations is generated for each control.  The above scheme therefore yields a balanced image data set, with 37,312 unique images belonging to each class.
  
  We reserve 10\% of the post-merger galaxies, and all of their associated images, and an equal number of control galaxies (2.5\%, since the control group contains four times as many galaxies), for testing. The remaining images are shuffled and split in traditional supervised learning fashion between the training set, which the networks study in detail, and the validation set, which is used to check progress from time to time, 90\% and 10\%, respectively. Different camera angles and mock observations of the same galaxy produce drastically different images. Therefore, training and validation data are partitioned by individual image and not by galaxy in order to maximize the morphological diversity contained in the training set, preparing the model for a broader range of possibilities.
  \smallskip
  \item \textit{Normalization}. Prior to training, all images are normalized such that the minimum pixel value is at zero, and the maximum pixel value is at one. We normalize images in linear fashion, by subtracting the value of the faintest pixel from the image, and then dividing by value of the brightest pixel. Since our CNN architecture requires images of a single size, we resize all images to 138 pixels on a side, while maintaining a constant physical width of 100 kpc. At $z$ = 0.102, the median simulated redshift of our galaxy population, an image with a physical scale of 100 kpc would be 138 CFIS pixels across. Consequently, a fixed size of 138 pixels allows us to minimize the total required amount of resizing. Galaxy images with mock-observed $z$ < 0.102 are downscaled, and those with $z$ > 0.102 are upscaled to meet this standard.
  \smallskip
  \item \textit{On-the-fly Augmentation}. Overfitting is of general concern for classification problems, and is of particular concern in this work because all but 10\% of the post-merger galaxies appear sixteen times in the combined training and validation sets. Each of the sixteen appearances are visually unique due to different redshifts and sky insertions. Hence, we rely on data augmentation to overcome this concern. Before each training epoch, we apply minor randomized image transformations using the ImageDataGenerator class in Keras (\citealp{chollet2015keras}) - vertical and horizontal shifts, shear transforms, and zooms of at most $\pm10\%$, as well as horizontal and vertical flips. In this way, even though the same source images are used in each round of training (epoch), the network never studies the same permutation of an image more than once.
  \end{itemize}

\subsection{CNN architecture}
\label{CNN Architecture}

We construct a 4-layer-deep CNN using a combination of open-source software tools in Python (Python Software Foundation. Python Language Reference, version 3.6\footnote{www.python.org/}), utilising Keras (\citealp{chollet2015keras}) for network construction, training, and data augmentation, and scikit-learn (Scikit-learn: Machine Learning in Python, \citealp{scikit-learn}) for partitioning the data.

Table \ref{arx-table} shows the constitutive Keras layers for the CNN, as well as the number of parameters associated with each. The general arrangement of the model is comparable to that of Alexnet (\citealp{NIPS2012_c399862d}). In all convolution layers, we use a stride of 1, and rectified linear unit activation function (\citealp{10.5555/3104322.3104425}). The hyper-parameters, including filter numbers, kernel sizes, and dropout percentages associated with each layer were optimized through an incremental iterative search.

\begin{table}
\begin{center}
\begin{tabular}{ |c|c|c|c| } 
\hline
Layer Type & \# Parameters & Output Shape\\
\hline
\hline
Input & 0 & (138,138,1) \\ 
\hline
\begin{tabular}{@{}@{}c@{}}Convolution \\ 32 Filters \\ Kernel (7,7)\end{tabular}  & 1600 & (138, 138, 32) \\ 
\hline
Max Pooling (2,2) & 0 & (69, 69, 32) \\ 
\hline
Dropout 25\% & 0 & (69, 69, 32)\\ 
\hline
\begin{tabular}{@{}@{}c@{}}Convolution \\ 64 filters \\ Kernel (7,7)\end{tabular} & 100416 & (69, 69, 64) \\ 
\hline
Max Pooling (2,2) & 0 & (34, 34, 64) \\ 
\hline
Dropout 20\% & 0 & (34, 34, 64) \\ 
\hline
Batch Normalization & 256 & (34, 34, 64) \\
\hline
\begin{tabular}{@{}@{}c@{}}Convolution \\ 128 filters \\ Kernel (7,7)\end{tabular} & 401536 & (34, 34, 128) \\ 
\hline
Max Pooling (2,2) & 0 & (17, 17, 128) \\ 
\hline
Dropout 20\% & 0 & (17, 17, 128) \\ 
\hline
\begin{tabular}{@{}@{}c@{}}Convolution \\ 128 filters \\ Kernel (7,7)\end{tabular} & 802944 & (17, 17, 128) \\ 
\hline
Max Pooling (2,2) & 0 & (8, 8, 128) \\ 
\hline
Dropout 20\% & 0 & (8, 8, 128) \\ 
\hline
\hline
Flatten & 0 & (8192) \\
\hline
Dense & 4194816 & (512) \\
\hline
Dropout 25\% & 0 & (512) \\
\hline
Dense & 65664 & (128) \\
\hline
Dropout 25\% & 0 & (128) \\
\hline
Activation, Sigmoid & 129 & (1) \\
\hline
\hline
\end{tabular}
\end{center}
\caption{The CNN architecture used in this work. Each layer begins with the stock Keras layer of the same name, with any specificed hyperparameters detailed in the Layer Type column. The \# Parameters column shows the number of trainable network parameters belonging to each layer.}
\label{arx-table}
\end{table}

A similar search was used for the optimization of training hyper-parameters, including the on-the-fly augmentation percentages detailed in Section~\ref{RealSimCFIS}, an Adadelta optimizer (\citealp{2012arXiv1212.5701Z}) with a learning rate of 0.05, and a training batch size of 32 images. The morphological diversity of both the post-merger and control classes introduces significant intrinsic inhomogeneity into the training data. Broadly, the final hyper-parameter recipe serves to combat overfitting to any particular sub-category within the data, and encourages the network to generalize.

\section{Results}
\label{Experiments}

Here we present the results of our experiments, first evaluating the CNN's performance by conventional machine learning metrics (Section~\ref{Test Set Properties}), and then by studying the stability of that performance as a function of simulation metadata and mock observation quantities (Section~\ref{Performance Trends}). We next produce a mock survey in the style of CFIS using galaxies from IllustrisTNG, and simulate an observational study of star formation enhancement (Section~\ref{Mock Survey}), before comparing our results to other automated methods (Section~\ref{Comparison to Automated Methods}), and finally to human classifications (Section~\ref{Visual Classification}).

\subsection{Overall CNN performance}
\label{Test Set Properties}

In order to evaluate our model's performance by traditional machine learning metrics, we apply the trained model to the reserved test galaxies, including 3,728 images each of post-mergers and controls, as detailed in Section~\ref{RealSimCFIS}. Because this data set does not contain any galaxies that the network has studied in training or referenced in validation, these results roughly correspond to the expected performance of the network should it be deployed to classify unseen data. In other words, one does not have to worry about the memorizing effects of deep neural networks (see also \citealp{2017arXiv170605394A}).

\begin{figure}
\includegraphics[width=\columnwidth]{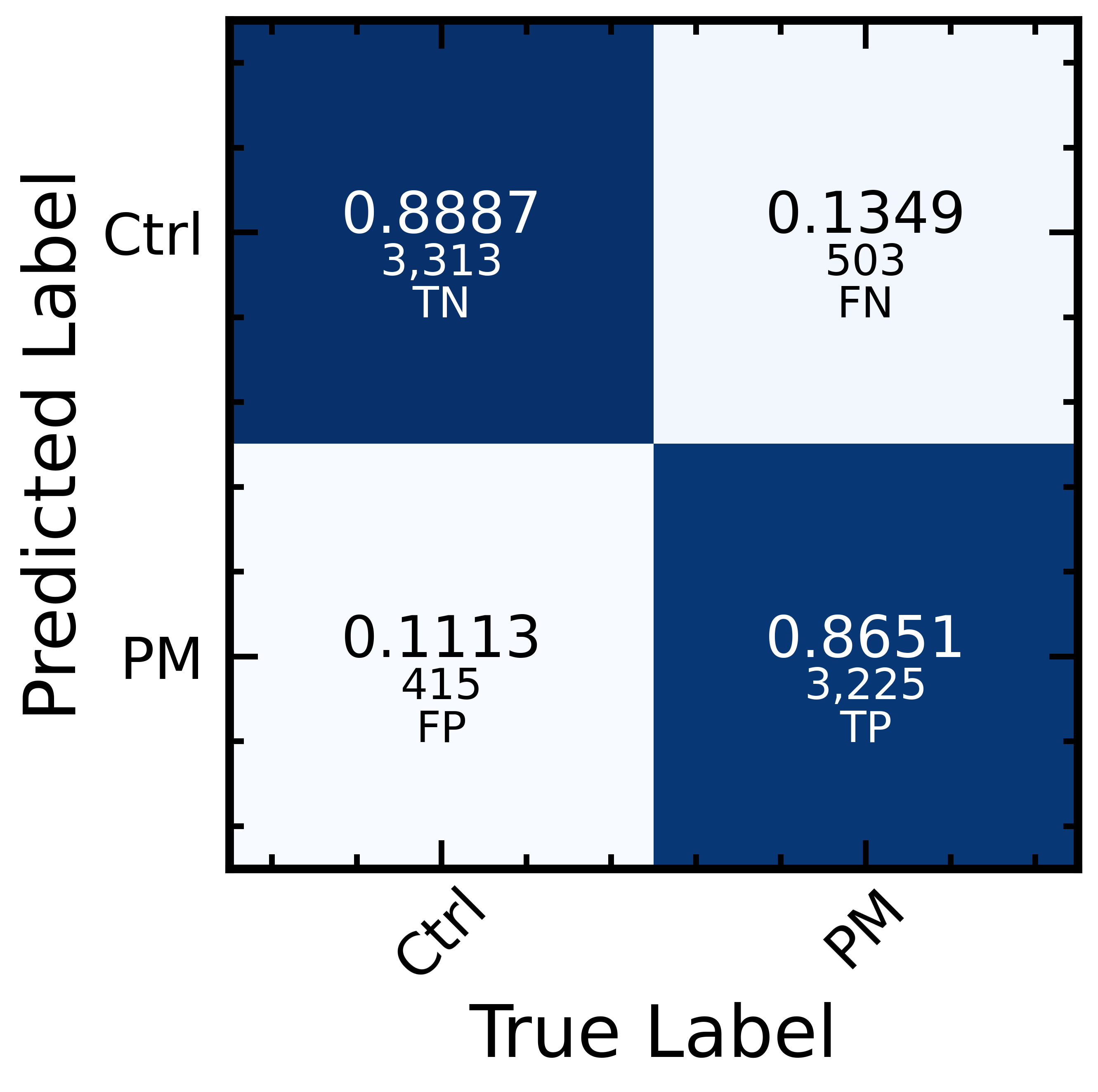}
\caption{Confusion matrix for the trained CNN applied to the never-seen images in the test set. Each quadrant is annotated with the normalized fractional accuracy, and the number of galaxies.  The total CNN performance (completeness) is 88\%.}
\label{fig:fs-test-cmx}
\end{figure}

The confusion matrix in Figure~\ref{fig:fs-test-cmx} shows the performance of the model evaluated on the reserved test galaxies. The model successfully identifies ~87\% of the post-merger images in the set, and ~89\% of the controls. Using training and test samples from the EAGLE cosmological simulations (\citealp{2015MNRAS.446..521S}), \citet{2019A&A...626A..49P} report 63\% and 67\% performance on mergers (including systems that are projected to merge in the next 0.3 Gyr) and non-mergers, respectively, after modulating their model's decision threshold to a position of 0.57. Also using TNG100, \citet{2020arXiv200902974W} report 76\% and 68\% for mergers and non-mergers, respectively, with the decision threshold at 0.53. However, the merger definition used by \citet{2020arXiv200902974W} also includes pre-mergers (galaxies that will undergo a merger in the next 1 Gyr), as well as post-mergers which have merged within 500 Myr. This comparison is therefore not exact.

\begin{figure}
\includegraphics[width=\columnwidth]{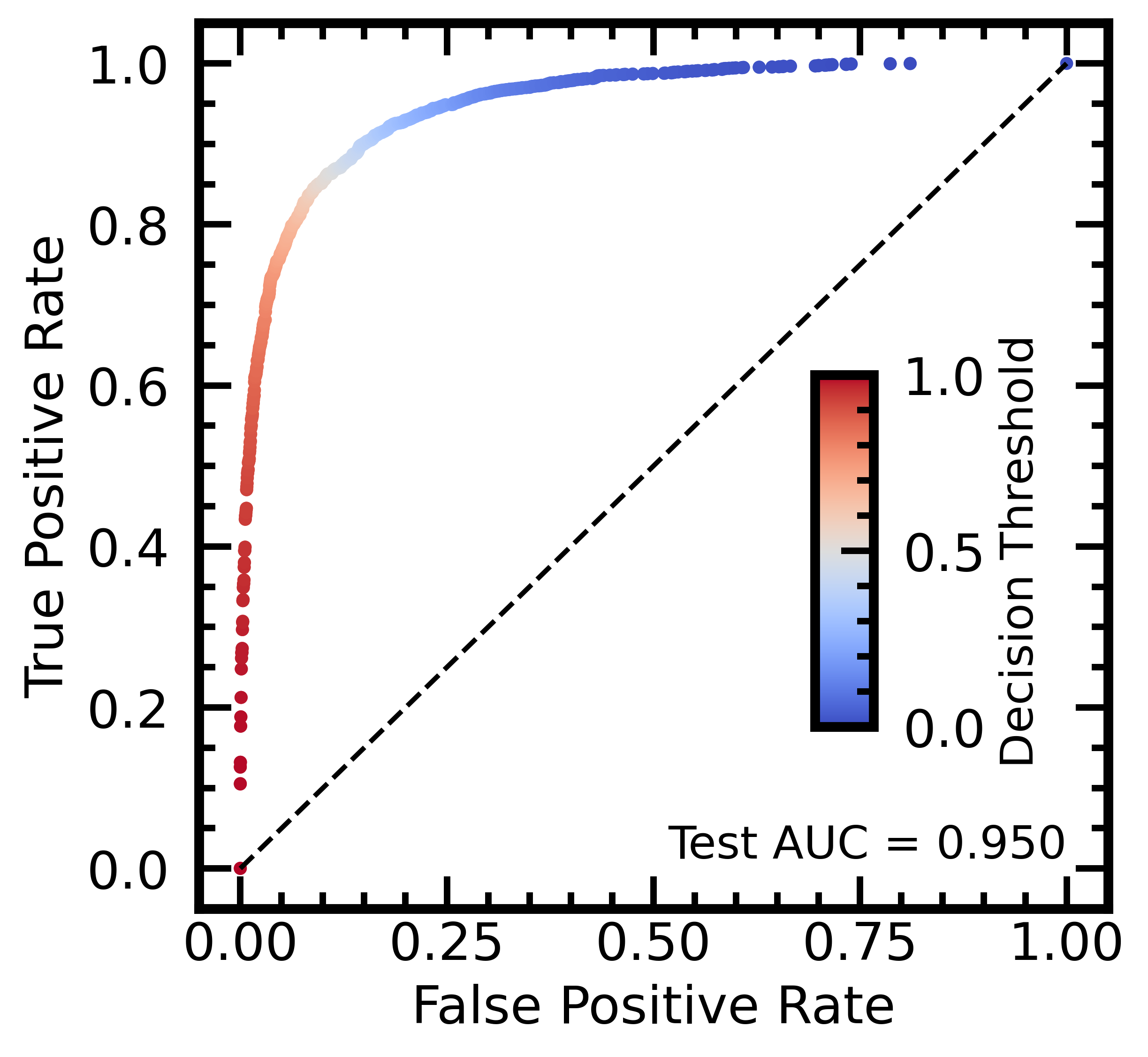}
\caption{The ROC curve with AUC score as the decision threshold is moved from 0 to 1 for the trained model as applied to the test set.}
\label{fig:fs-test-roc}
\end{figure}

The Receiver Operating Characteristic (ROC) curve in Figure~\ref{fig:fs-test-roc} shows the false positive rate (i.e. the fraction of controls that are mislabeled as post-mergers) and the true positive rate (i.e. the fraction of post-mergers that are correctly identified) as the model's decision threshold is modulated from zero (label everything as a post-merger) to one (label everything as a control). The default threshold for classification is 0.5. The choice of decision threshold is examined later in Section~\ref{Mock Survey}. The dashed diagonal line is characteristic of a hypothetical model with random label assignment by the model. The area under the model's performance curve (AUC) serves as a conventional metric of classification performance, with a value of 1.0 indicating perfect performance; networks that develop a strong grasp of the classification problem at hand are likely to have high AUC scores. Promisingly, our trained model achieves an AUC score of 0.95.

\subsection{CNN performance trends}
\label{Performance Trends}

Having assessed the CNN's overall performance on the test set, we next study its dependence on the galaxies' intrinsic (e.g. stellar mass, gas fraction and redshift) and environmental properties (e.g. near neighbour proximity). To this end, we re-sample the post-mergers and the first round of controls from a new, fifth camera angle, at a vertex in the first octant of a cube with the galaxy at its centre. The resulting stellar mass maps are re-processed with randomized \textsc{RealSim-CFIS} parameters, assigning new sky locations, observational noise, and mock observation redshift values on an image-wise basis. We also re-train the CNN, folding the previously-reserved test galaxies into the training pool, shuffling on an image-wise basis, and partitioning them into test (90\%) and validation (10\%) data. Allowing the model to study the entire galaxy sample as imaged from the first four camera angles eliminates any image memorization bias when evaluating its performance on the resampled data. The CNN's global performance on the resampled data is the same, 88\% accuracy, implying that the new projected morphology and mock observation are sufficient to prevent identification via memorization on the galaxies studied in training. While the reserved test galaxies from Section~\ref{Test Set Properties} are chosen at random, and therefore represent a equitable sampling of the full population, re-introducing galaxies previously reserved for training and validation improves the significance of our statistics and drives down the scatter of our performance metrics, particularly in regions of a given parameter that are sparsely populated.

After obtaining model predictions for the resampled data, we bin the constituent galaxies, as well as the model's predictions on them, by a selection of IllustrisTNG metadata quantities. This allows us to study the natural biases imparted on the model by the visual characteristics of the galaxies. Figures~\ref{fig:fs-test-perf-vs-r1}-\ref{fig:fs-test-perf-vs-gf} show the results of these tests and have the same format throughout. Top panels show the raw number of post-mergers and controls as a function of the property under investigation (e.g. redshift, stellar mass etc.). These numbers are relevant for performance assessment as we might expect the model to fare more poorly in regimes where it has been exposed to fewer examples in training. The histograms are further divided to show the number of galaxies that are correctly or incorrectly classified, distinguishing between true positives (tp; post-mergers correctly classified as such), false positives (fp; controls that are erroneously classified as post-mergers), true negatives (tn; controls correctly classified as such) and false negatives (fn; post-mergers that are erroneously classified as controls). In the lower panels of Figures~\ref{fig:fs-test-perf-vs-r1}-\ref{fig:fs-test-perf-vs-gf} we show the fraction of controls and post-mergers that are correctly classified. The horizontal dashed line shows the model's average performance for reference. In all figures, the \textit{x} error bars show the width of the bins and the \textit{y} error bars are the binomial errors in each bin, $\sqrt{f*(1-f)/N}$ where f is the fraction correctly identified, and N is the number of galaxies in the bin. The pink curve shows the model's average performance on all galaxies in the bin, the blue curve shows the specific performance on post-mergers, and the orange shows the specific control (non-post-merger) performance. In the following sub-sections we describe in more detail the results and conclusions from Figures~\ref{fig:fs-test-perf-vs-r1}-\ref{fig:fs-test-perf-vs-gf}.

\subsubsection{Role of environment}
\label{Role of Environment}

\begin{figure}
\includegraphics[width=\columnwidth]{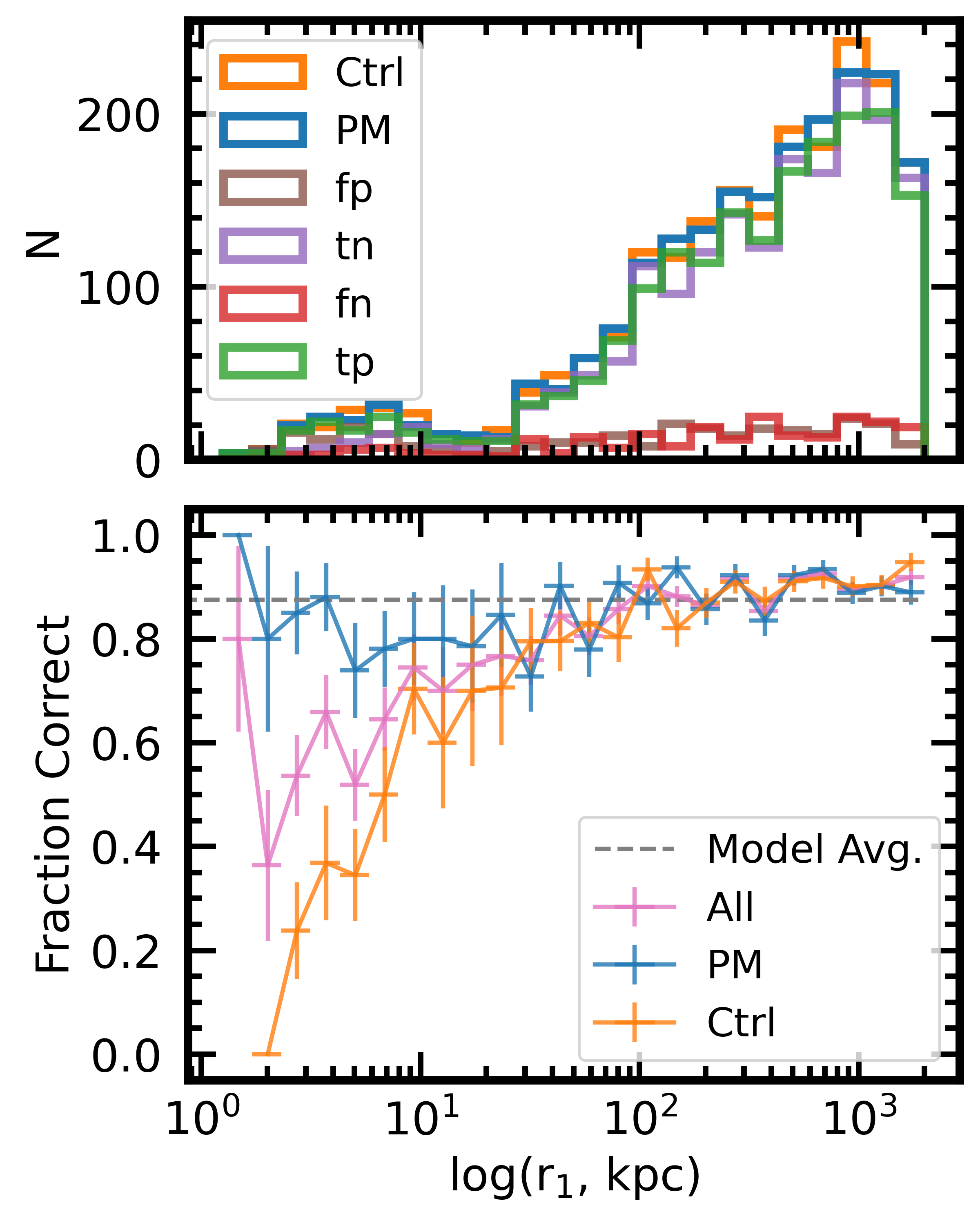}
\caption{The trained model's performance, as a function of separation to the nearest neighbour ($\mathrm{r_{1}}$). The top panel shows the raw number of post-mergers and controls (blue and orange histograms, respectively), further broken down as correctly and incorrectly classified (fp, brown: controls classified as post-mergers; tn, purple: correctly-classified controls; fn, red: post-mergers classified as controls; tp, green: correctly-classified post-mergers). The bottom panel shows the fraction of post-merger and control galaxy images correctly identified by the model. Horizontal error bars for all points are the bin widths, and vertical error bars are the binomial errors for each bin, $\sqrt{f*(1-f)/N}$ where f is the fraction correctly identified, and N is the number of galaxies in the bin.}
\label{fig:fs-test-perf-vs-r1}
\end{figure}

We find that a galaxy's simulated environment can strongly influence how it will be classified by a CNN model. Indeed, for networks whose aim is to identify galaxy pairs (e.g. \citealp{2018MNRAS.479..415A}, \citealp{2019MNRAS.490.5390B}) the presence of a close companion is an essential piece of information. In order to test whether our post-merger classification is affected by the presence of a close companion, we investigate in Figure~\ref{fig:fs-test-perf-vs-r1} the fraction of correctly classified galaxies as a function of separation to the nearest neighbour ($\mathrm{r_{1}}$). Figure~\ref{fig:fs-test-perf-vs-r1} shows that when a neighbour galaxy appears within  \textasciitilde50 kpc (the radial extent of our images), the model grows uncertain, assigning post-merger classifications to numerous control galaxies, and yielding below-average total performance. Figure~\ref{fig:fs-test-perf-vs-r1} demonstrates that the model retains much of its ability to distinguish between post-mergers and non-post-mergers with a potential pre-coalescence partner down to 10 kpc, below which the visual degeneracy becomes prohibitive. While close pairs are, for our purposes, not necessarily post-mergers, and may be counted as misclassified by the model, in many cases the CNN may still be identifying genuine merger features in pre-coalescence pairs. Still, such close neighbours are rare in both the simulation and the real Universe, and hence should not present a significant source of contamination to the out-falling post-merger sample.

\subsubsection{Role of observed redshift}
\label{Role of observed redshift}

\begin{figure}
\includegraphics[width=\columnwidth]{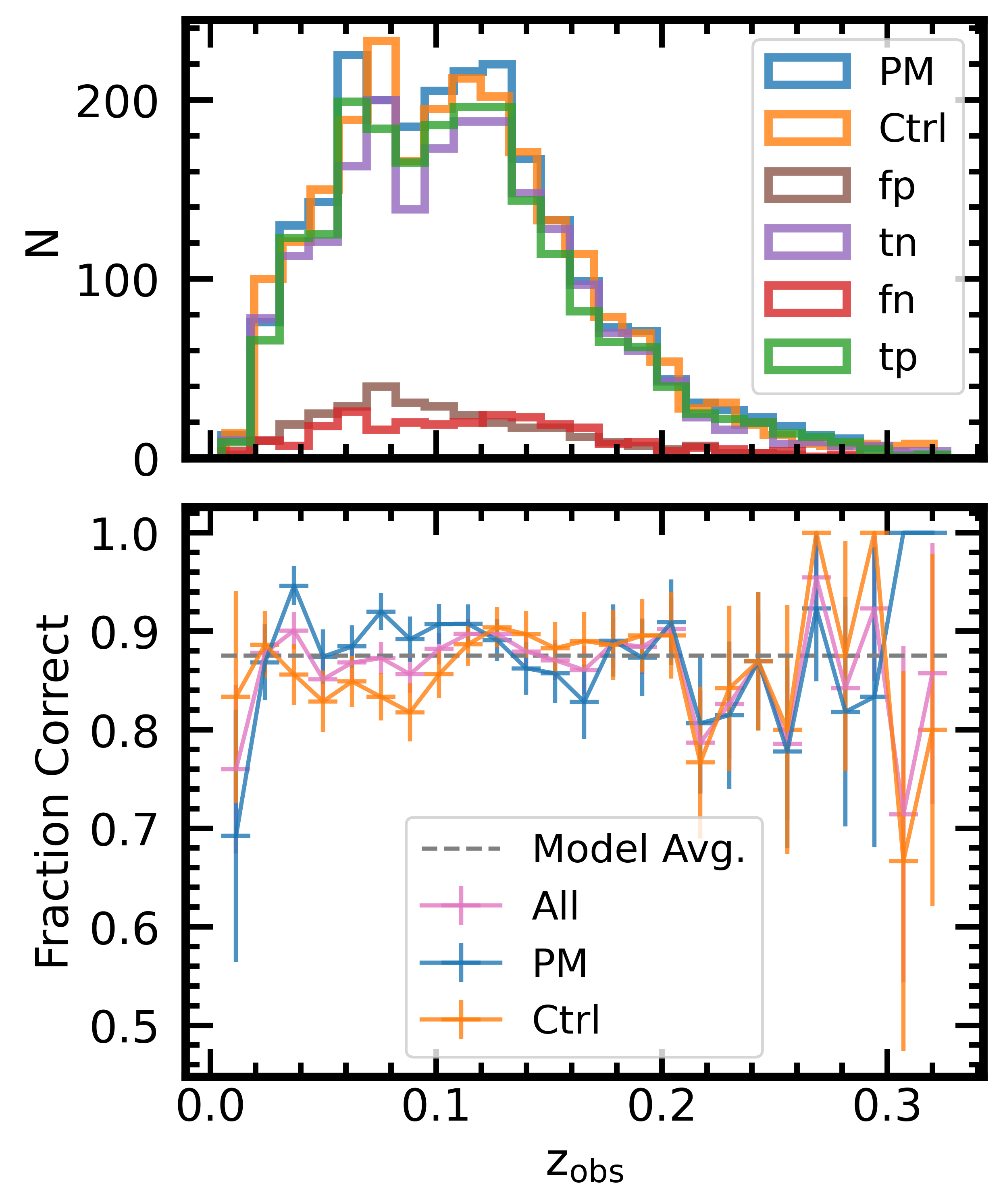}
\caption{As for Figure~\ref{fig:fs-test-perf-vs-r1} but performance is plotted as a function of mock observed redshift.}
\label{fig:fs-test-perf-vs-z}
\end{figure}

Figure~\ref{fig:fs-test-perf-vs-z} shows the performance of the trained model as a function of insertion redshift. We remind the reader that this is selected at random from the observed SDSS distribution, and is not linked to the simulation redshift. It is perhaps natural to expect that performance may dwindle at higher redshifts, due to loss of spatial resolution and the dimming of faint features. For insertion redshift values of \textasciitilde0-0.2, where galaxies are better-resolved and brighter relative to the sky, performance is consistent with the overall model average, with a slight enhancement in post-merger identification in the lowest mock observation redshift bins. Above $z$ = 0.2, degrees of uncertainty are introduced as the resolution diminishes and low-surface brightness merger features grow indistinguishable from the background noise. A further contributing factor may also be the relative paucity of training images at $z$ > 0.2, as these are rarer in the redshift distribution from which the mock images are generated. We find that artificially enhancing the number of training images with low mock observed redshifts, or indeed, in any sparse tract of parameter space, improves the CNN's performance there. However, with a finite number of training examples, this improved stability comes at the cost of global performance. Using realistic parameter distributions budgets the training material optimally, and prepares the CNN to identify galaxies as they are likely to appear in a natural sample.

\subsubsection{Role of galaxy mass}
\label{Role of galaxy mass}

\begin{figure}
\includegraphics[width=\columnwidth]{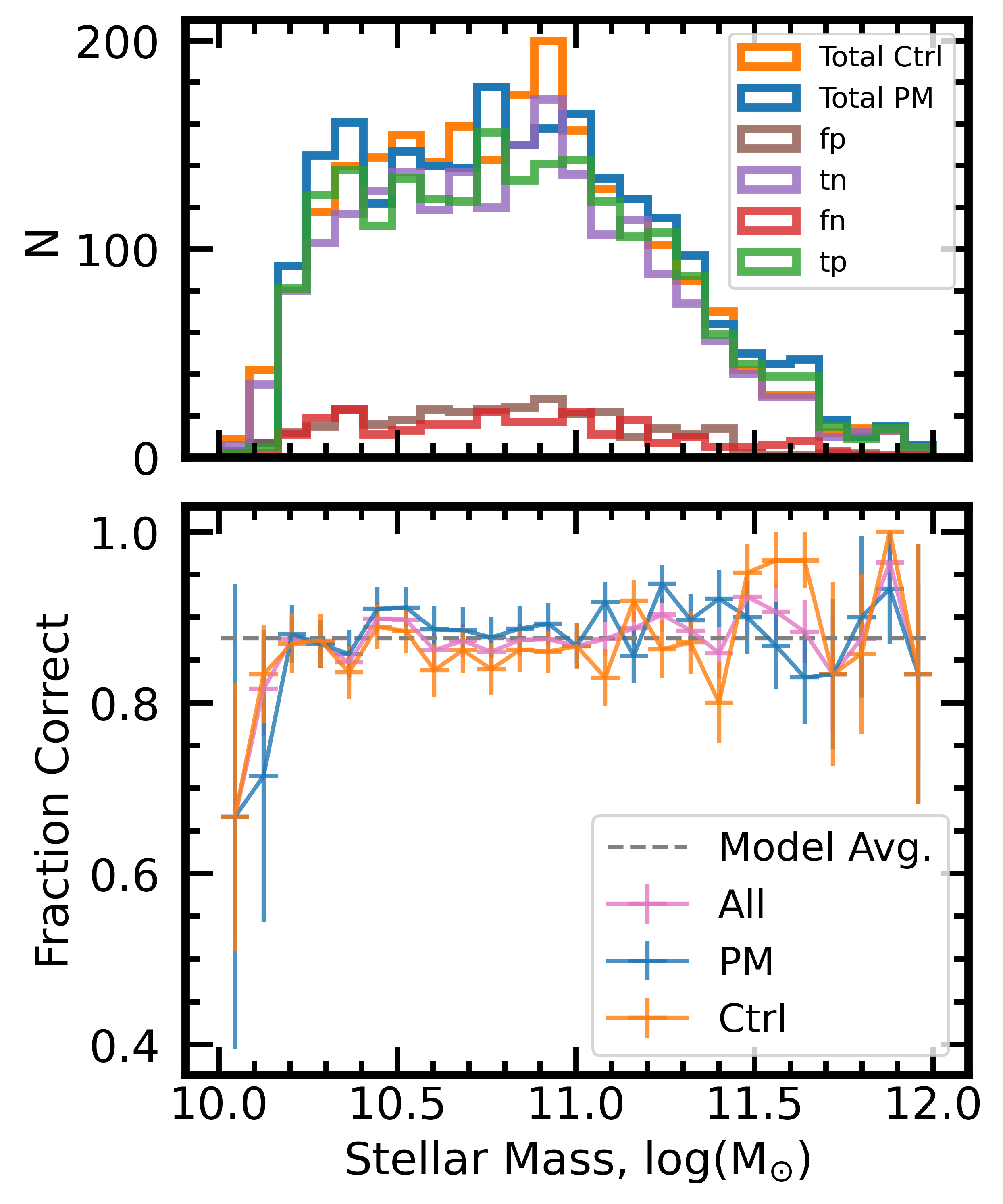}
\caption{As for Figure~\ref{fig:fs-test-perf-vs-r1} but performance is plotted as a function of galaxy stellar mass.}
\label{fig:fs-test-perf-vs-mstar}
\end{figure}

Performance with stellar mass (Figure~\ref{fig:fs-test-perf-vs-mstar}) is consistent with the model's average where the bulk of the selected galaxies (\textasciitilde$\mathrm{10}^\mathrm{10.2}$-$\mathrm{10}^\mathrm{11.5}$ \(\textup{M}_\odot\)) are found. Performance drops for the lowest stellar mass bins, and becomes unstable in higher stellar mass bins, where fewer galaxies of both classes exist for the network to study in the training phase. The parent galaxy luminosity function can be modeled by a power law in the low-mass regime, and so a higher number of low-mass galaxies might typically be expected. While they may be numerous, low-mass galaxies in hydrodynamical simulations are also more likely to be impacted by numerical stripping, an effect where stellar mass particles in the outskirts of a low-mass galaxy may erroneously be assigned to a nearby galaxy of greater mass (e.g. \citealp{2015MNRAS.449...49R}). The abrupt shelf on the left-hand side of the top-panel histograms in Figure~\ref{fig:fs-test-perf-vs-mstar} can therefore be attributed to our selection criteria, which were developed to minimize the effects of numerical stripping in the sample (see also \citealp{2020MNRAS.493.3716H}).

\subsubsection{Role of merger mass ratio}
\label{Role of merger mass ratio}

\begin{figure}
\includegraphics[width=\columnwidth]{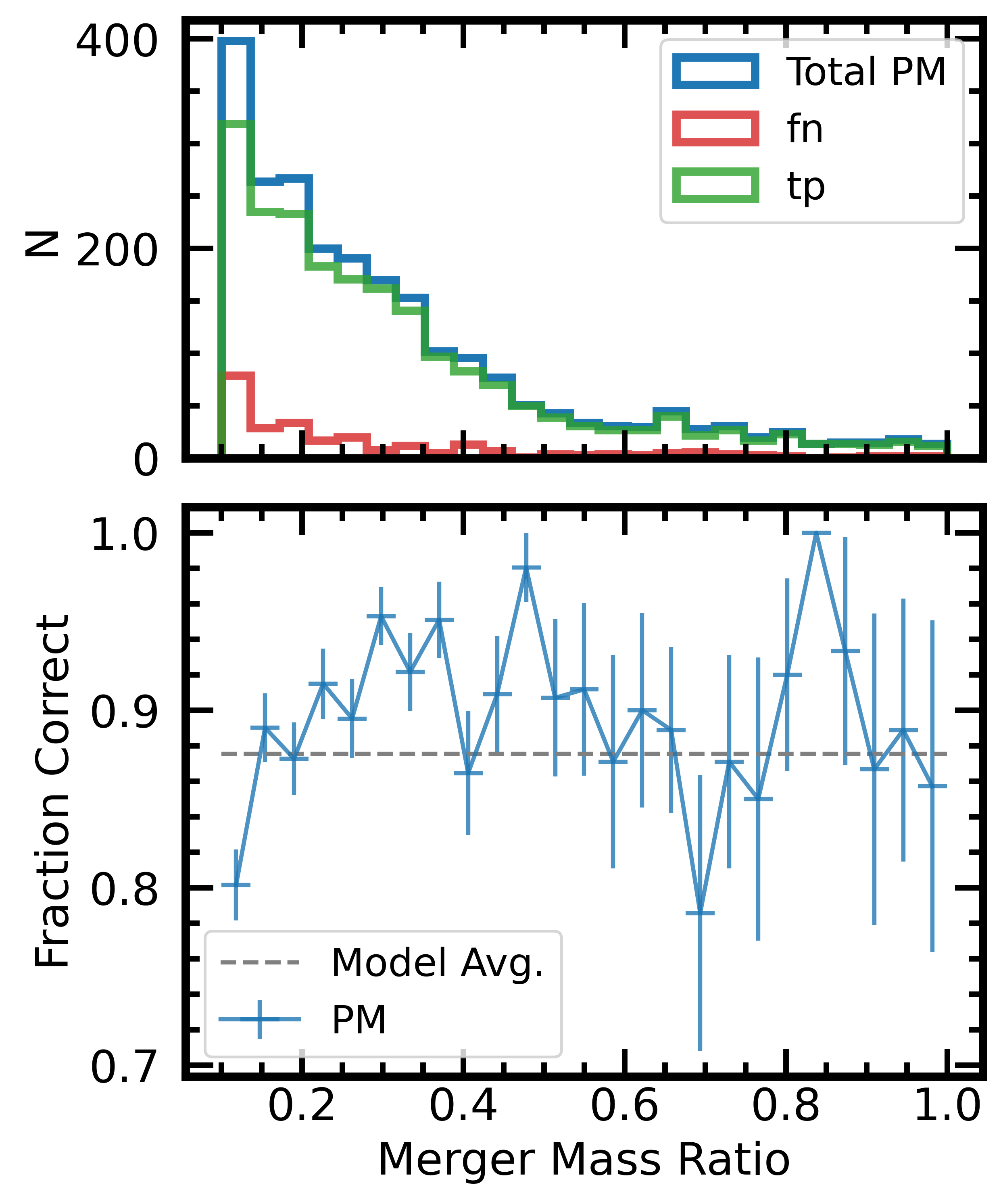}
\caption{As for Figure~\ref{fig:fs-test-perf-vs-r1} but performance is plotted as a function of merger mass ratio. Post-mergers used in training must have a mass ratio of $\geq$ 0.1.}
\label{fig:fs-test-perf-vs-mr}
\end{figure}

Figure~\ref{fig:fs-test-perf-vs-mr} shows the model's performance as a function of the ratio of the smaller participant galaxy's stellar mass to that of the larger. The model exhibits minimal volatility in classifying mergers across different mass ratios, with a minor suppression in identifying mergers that are comparatively minor (with a mass ratio of < 0.2), as well as slight instability for mergers near 1:1, of which there are very few examples to study (see blue histogram, top panel of Figure~\ref{fig:fs-test-perf-vs-mr}). The CNN offers nearly perfect performance for moderate to major mergers (\textasciitilde0.1-0.6), with minimal scatter in high mass ratio bins with a small number of objects. Because the simulation snapshot timescale is only \textasciitilde160 Myr, we can compare to other merger mass ratio studies with the assumption that the galaxies in our post-merger sample have an effective $\mathrm{T}_\mathrm{Postmerger}$ of zero. We do not uncover a strong link between mass ratio and classification performance as in \citet{2019ApJ...872...76N}, but our finding is consistent with the \citet{2010MNRAS.404..575L} result that post-merger observability within 0.2-0.4 Gyr is largely insensitive to mass ratio.

\subsubsection{Role of gas reservoir}
\label{Role of gas reservoir}

\begin{figure}
\includegraphics[width=\columnwidth]{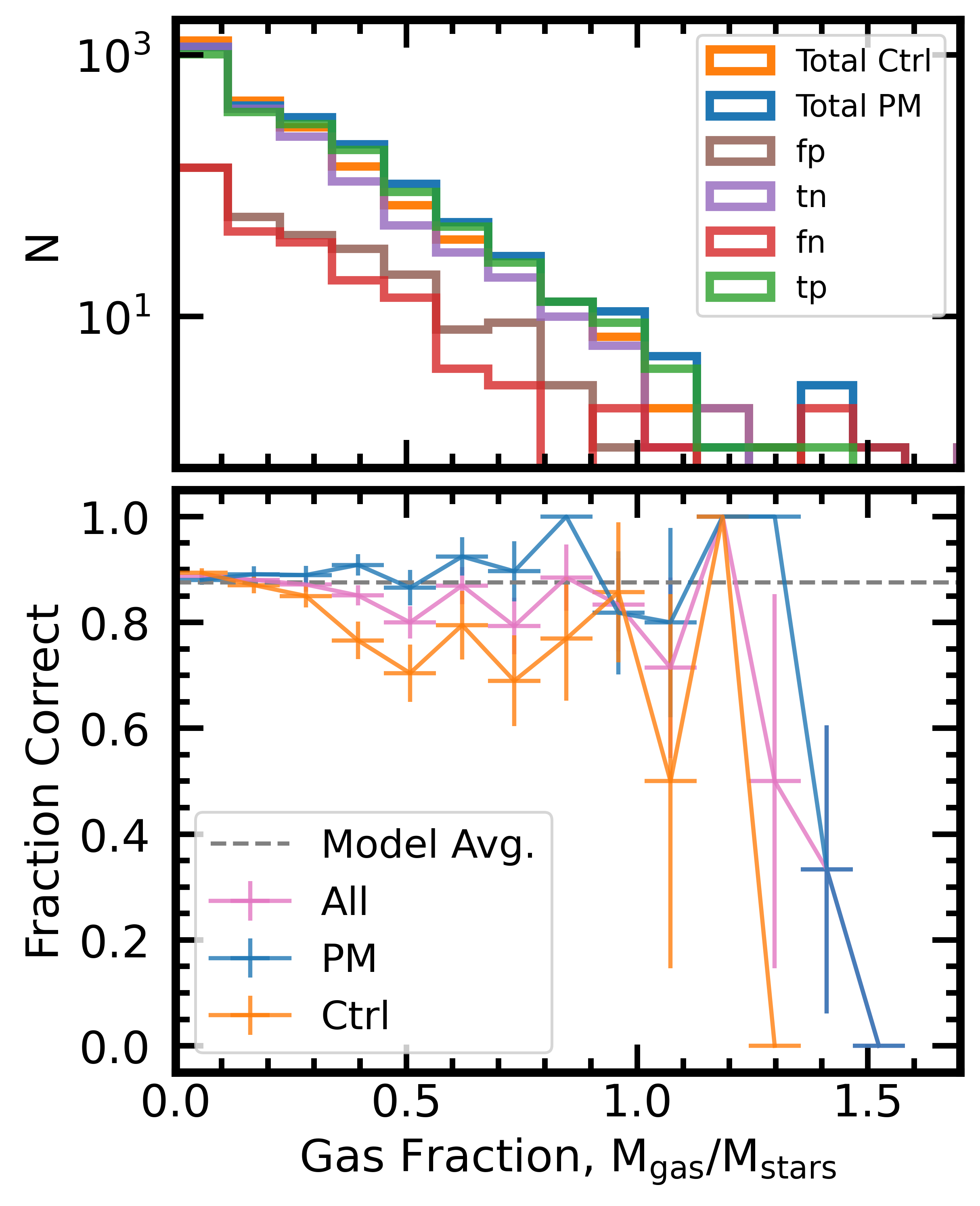}
\caption{As for Figure~\ref{fig:fs-test-perf-vs-r1} but performance is plotted as a function of the post-coalescence gas fraction. Although the images on which the model is trained contain no gas information, the strong trend is a secondary effect with stellar mass.}
\label{fig:fs-test-perf-vs-gf}
\end{figure}

Simulations have previously indicated that gas fraction can affect the observability of the merger phase (e.g. \citealp{2010MNRAS.404..590L}; \citealp{2014A&A...566A..97J}). Figure~\ref{fig:fs-test-perf-vs-gf} tests the model's performance as a function of gas fraction. We note that this is the gas fraction of the post-merger remnant and not the incoming galaxies, although this metric should still broadly capture whether (at least one of) the merging galaxies had significant gas reservoirs. Figure~\ref{fig:fs-test-perf-vs-gf} shows that in the regime where most of the post-mergers are located (gas fractions less than unity), post-mergers are consistently well classified, even at low gas fractions. Conversely, control galaxies are increasingly poorly classified towards higher gas fractions. Because synthetic observations are generated without any gas information, we might not expect to uncover a dramatic performance trend with gas fraction. However, we characterize this trend as a secondary effect with stellar mass: nearly all galaxies with gas fractions $\mathrm{M_{Gas}/M_{\star}}$ > 0.3 belong to the lowest bin of stellar mass as seen in Figure~\ref{fig:fs-test-perf-vs-mstar}. By design, most galaxies in the sample have moderate stellar masses and typical gas fractions, and so this apparent dive in classification accuracy is insufficient to diminish performance at large.

\subsection{Mock survey}
\label{Mock Survey}

In a matched galaxy sample, like those studied by the CNN in Sections~\ref{Test Set Properties} and~\ref{Performance Trends}, the enforced balance of post-mergers and controls allows the model to enjoy superficially high purity percentages, even with minor adjustments to the classification decision threshold. However, the value of any automated system designed to identify post-merger galaxies lies in its ability to do so within the framework of a large observational sample. One of the most striking distinctions between such a sample and the image data we have prepared up to this point is the enforced over-abundance of post-mergers relative to non-post-mergers. However, mergers are expected to represent only a few percent of galaxies in the observable (\citealp{1993MNRAS.262..627L}; \citealp{2011ApJ...742..103L}; \citealp{2012ApJ...747...34B}; \citealp{2014MNRAS.445.1157C}), and simulated (\citealp{2015MNRAS.449...49R}; \citealp{2018MNRAS.480.2266M}) low-redshift universes. Moreover, the ultimate application of our trained model is the identification of a pure sample of post-mergers in CFIS that will allow us to study the properties of galaxies after their coalescence.

\begin{figure}
\includegraphics[width=\columnwidth]{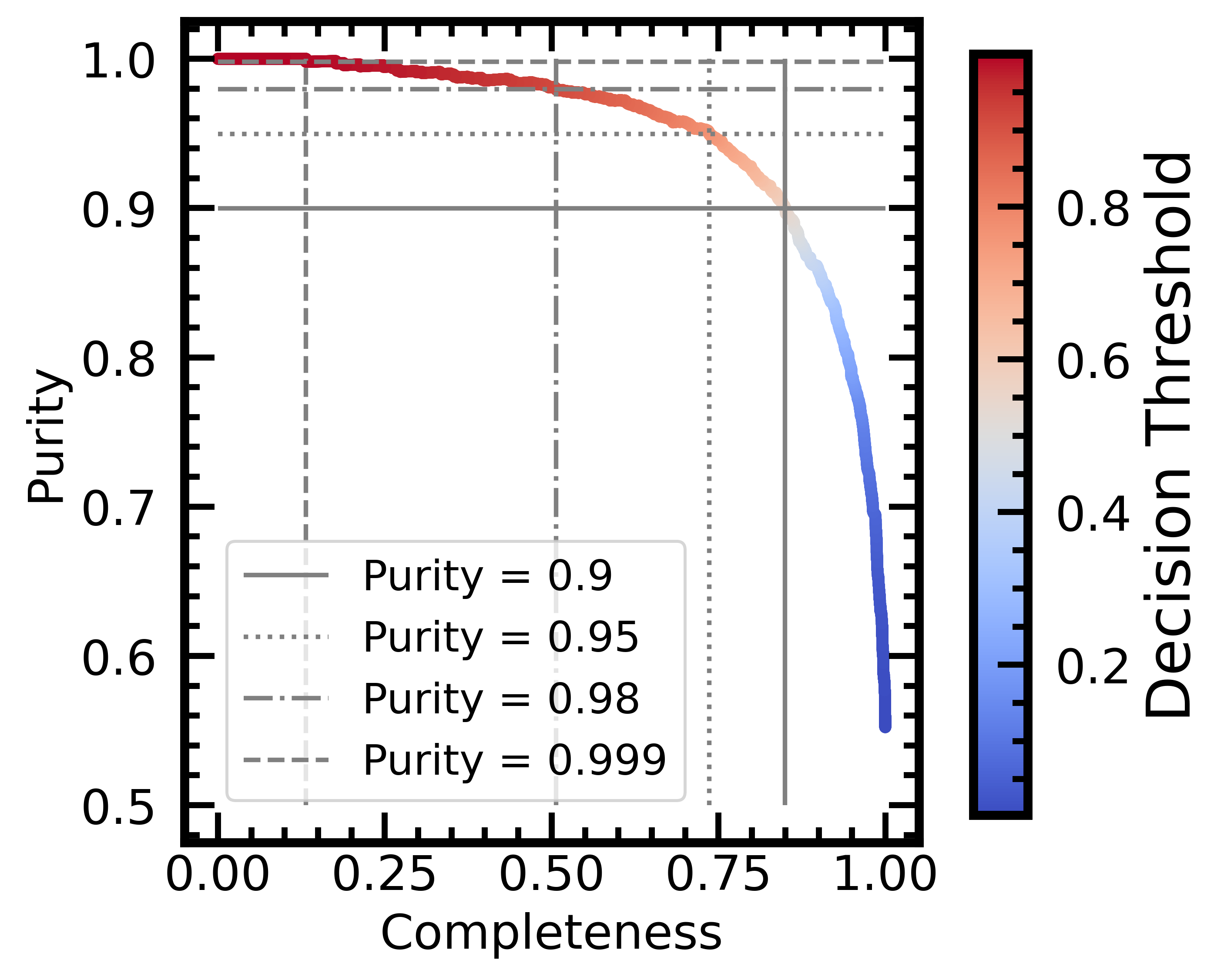}
\caption{The purity-completeness (or precision-recall) curve for the CNN's performance on the test set, which has equal numbers of post-merger and control galaxies. Annotations show the decision threshold setting and completeness cost required to achieve various pre-specified sample purity values.}
\label{fig:exp-prc}
\end{figure}

Figure~\ref{fig:exp-prc} shows the trade-off between purity and completeness (alternatively known as precision and recall) of the out-falling post-merger sample in the test set. The default threshold of 0.5 yields a purity of 89\% in a balanced data set. As an example of the impact of this threshold in a survey with less than 1\% actual post-mergers, the expected purity of the resulting post-merger sample (from Bayes theorem, discussed further in Section~\ref{Limitations}) would be only \textasciitilde6\%. In order to explore the impact of the combination of the natural scarcity of mergers with the imperfect identification from any automated (or, indeed, visual) classification method, we prepare and study a mock survey, with post-mergers and non-post-mergers present in quantities representative of TNG100-1.

\subsubsection{Survey data preparation}
\label{Survey Data Preparation}
Following the approach of Section~\ref{Performance Trends}, we perform a single synthetic observation of every galaxy in TNG 100-1 with a stellar mass of $\mathrm{M_{\star}}$ > $\mathrm{10}^\mathrm{10}$ \(\textup{M}_\odot\) from the fifth camera angle, at a vertex in the first octant of a cube with the galaxy at its centre. Since galaxies are randomly oriented with respect to the simulation box, this camera angle is consistent with a random set of orientations, while also projecting different apparent morphology than would have appeared in the original four camera angles used in training. Since this complete data set does not contain matched pairs of post-mergers and controls, individual mock observation redshift values are selected at random for each object on the fly. In total, the resulting mock survey contains one image each for the 303,110 galaxies. 2332 (0.7\%) of the images are of post-merger galaxies in their first post-coalescence snapshot, while the rest, all galaxies with  $\mathrm{T}_\mathrm{Postmerger}$ > 0, are counted as non-post-mergers for the purposes of this experiment. The non-post-merger category in the mock survey is therefore distinct from the control group used up to this point.

\subsubsection{Survey training modifications}
\label{Survey Training Modifications}
Evaluating on a set of reserved test galaxies in Section~\ref{Test Set Properties}, as well as on a larger resampled galaxy population in Section~\ref{Performance Trends}, established that our combination of the training data, network architecture, and training hyper-parameters were sufficient to grasp the nuances of the classification problem without depending on memorization of specific images. For the mock survey, therefore, we will use the same model as in Section~\ref{Performance Trends}, which has studied the entire matched galaxy sample from the first four camera angles, and present its performance on the 303,110-image data set detailed in Section~\ref{Survey Data Preparation}.

\subsubsection{Mock survey classification}
\label{Mock Survey Classification}

\begin{figure}
\includegraphics[width=\columnwidth]{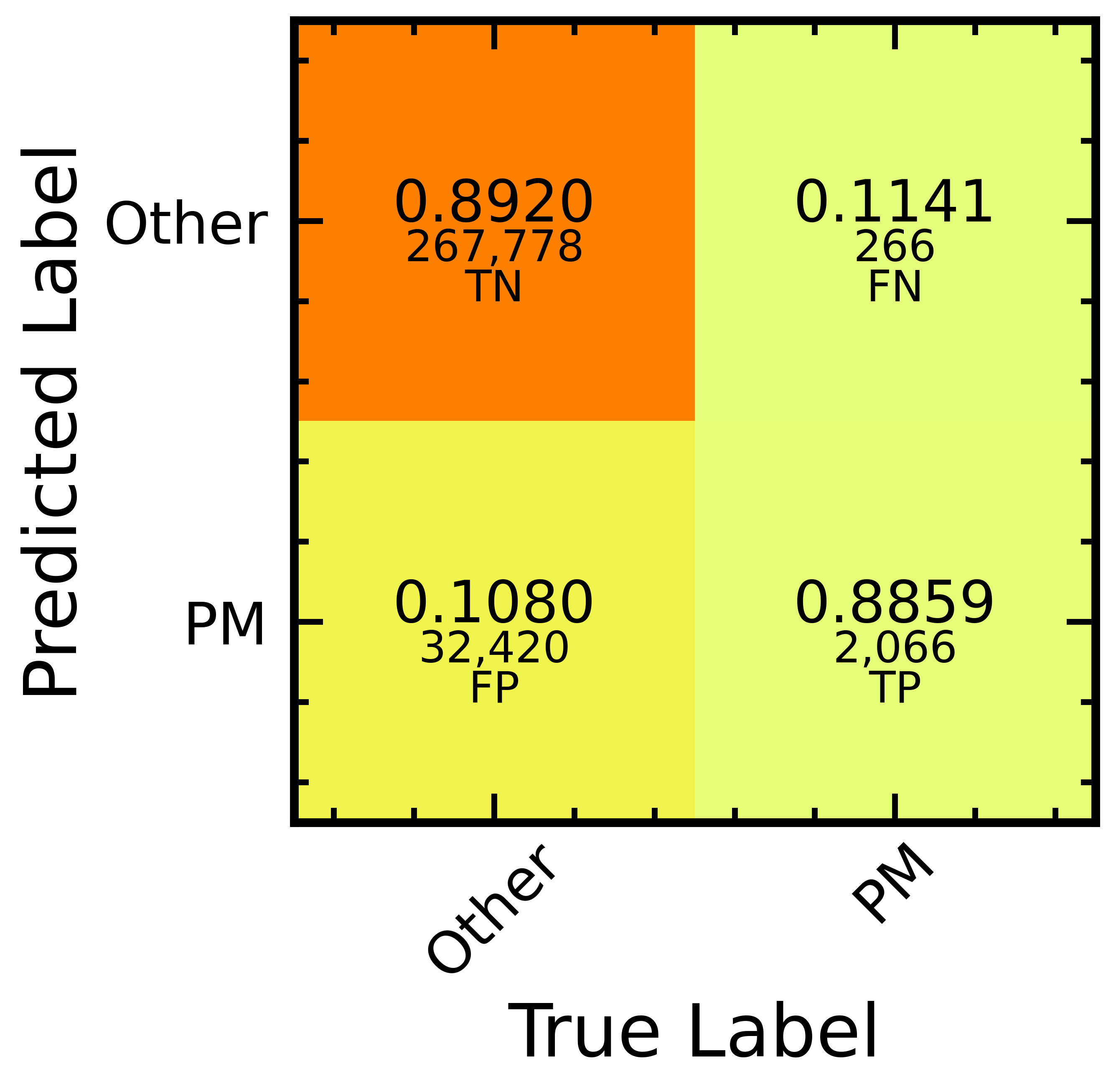}
\caption{Confusion matrix for the fully-trained model's performance on the mock survey images. Each quadrant is annotated with the normalized fractional accuracy, and the number of galaxies.  Although the overall sample completeness is 89\% for post-mergers and controls, the scarcity of post-mergers leads to a relatively small sample.}
\label{fig:mock-surv-cmx}
\end{figure}

Figure~\ref{fig:mock-surv-cmx} shows the mock survey confusion matrix for the model trained on all of the images in the original matched data set, with no reserved test galaxies. Pure morphological memorization is unlikely to work reliably, due to the use of a new fifth camera angle in generating the mock survey. Still, the model successfully identifies \textasciitilde89\% of both the mergers in their first post-coalescence snapshot, and the non-post-mergers, a category which now includes any object of mass >$\mathrm{10}^\mathrm{10}$ \(\textup{M}_\odot\) not explicitly belonging to the post-merger category. Despite strong fractional completeness of both classes and an AUC score of over 95\%, non-post-mergers in the simulation outnumber post-mergers by more than 100 to 1, and so the number of false positives (non-post-mergers identified as post-mergers by the model) outnumber the true positives (correctly labeled post-mergers) almost 14 to 1, resulting in an out-falling post-merger purity of only \textasciitilde6\% (a quantity in good agreement with that predicted by Bayes rule, \citealp{1763RSPT...53..370B}). Such a low a purity is obviously problematic if the eventual science goal is to assess the statistical properties of the CNN-identified post-merger sample (e.g. \citealp{2019A&A...626A..49P}). In spite of the disadvantage, however, the model distills the mock survey efficiently, returning a predicted post-merger set that contains $\frac{1}{10}$ as many images as the mock survey. Only 0.7\% of the mock survey galaxies are post-mergers with $\mathrm{T}_\mathrm{Postmerger}=0$ Gyr, but the CNN's predicted post-merger sample is \textasciitilde9 times as distilled (i.e. the true post-merger fraction in the sample is greater by a factor of \textasciitilde9, up to 6\%). The predicted post-merger sample is also \textasciitilde10 times as distilled (10\% from 1\%) in post-mergers with $\mathrm{T}_\mathrm{Postmerger}$ values of $\le0.2$ Gyr. This suggests that the CNN continues to identify legitimate post-merger features for hundreds of Myr after the merger occurs (see also Section~\ref{Role of time-since-merger}). As in Section~\ref{Role of Environment}, we find that a number of non-post-merger galaxies with a nearby neighbor still contaminate our post-merger sample. However, many of these misclassifications are also due to the presence of merger-related features. 46\% of false positive galaxies in the mock survey go on to experience a merger within 500 Myr, while only 33\% do not experience a merger in the next 2 Gyr. Therefore, while there is a meaningful amount of genuine contamination, the network shows a clear preference towards pre-mergers with merger-induced features compared to galaxies that experience a flyby.

\begin{figure}
\includegraphics[width=\columnwidth]{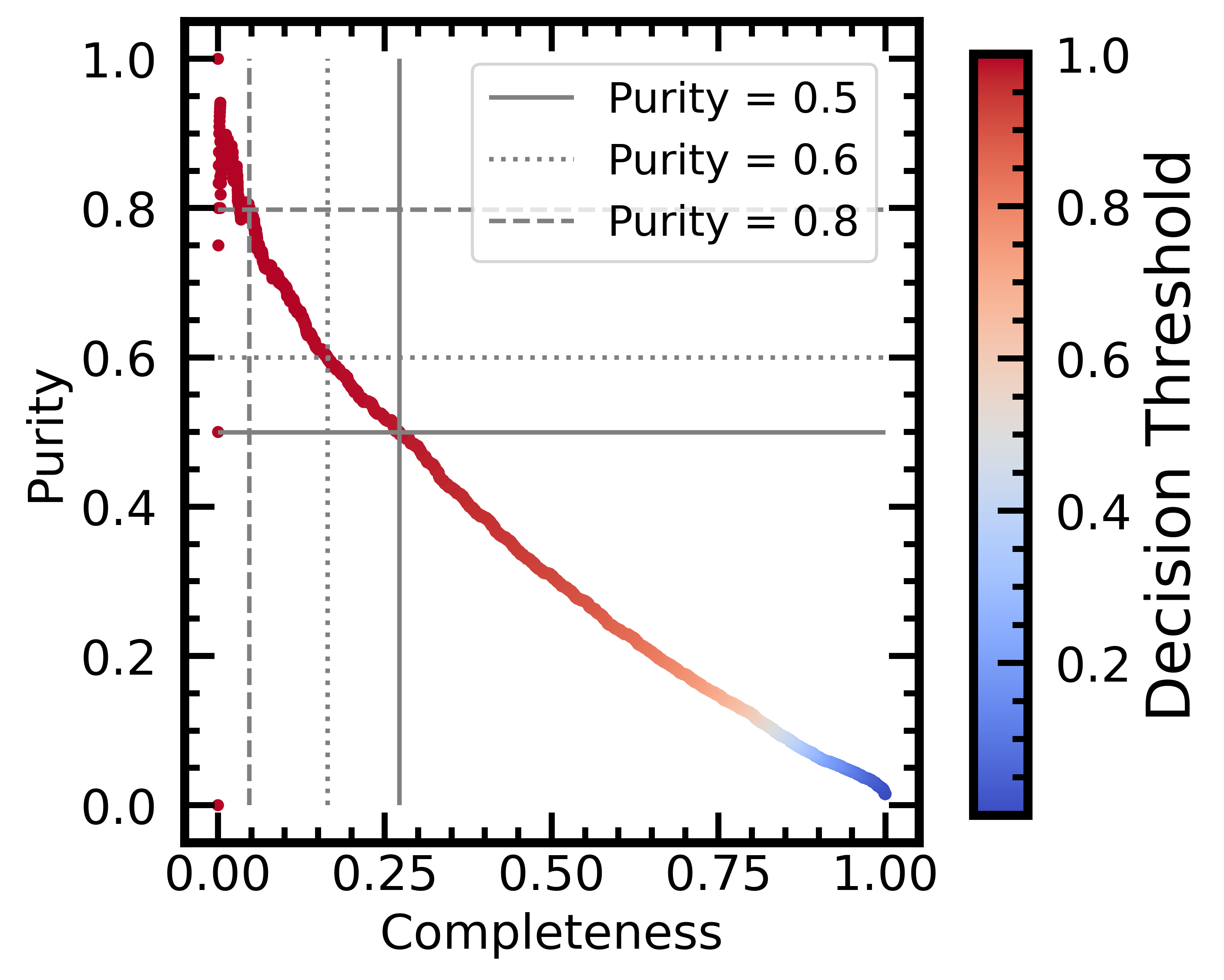}
\caption{The purity-completeness (or precision-recall) curve for the mock survey. For the default decision threshold at 0.5, the purity is very low, approximately 6\%. Increasing the threshold as specified in the annotation allows for higher sample purity to be achieved (see also Table~\ref{mock-surv-prc-table}).}
\label{fig:mock-surv-prc}
\end{figure}

\begin{table}
\begin{center}
\begin{tabular}{ |c|c|c|c| } 
\hline
Threshold & \ Purity & \ Completeness & \ PM Sample Size  \\
\hline
\hline
0.5 & 0.06 & 0.89 & 34486 \\
\hline
0.981 & 0.5 & 0.27 & 2535 \\
\hline
0.992 & 0.6 & 0.16 & 1531 \\
\hline
0.999 & 0.8 & 0.05 & 439 \\
\hline
\hline
\end{tabular}
\end{center}
\caption{Decision thresholds required to yield example purity levels in the mock survey, which contains 303,110 galaxies and 2332 actual post-mergers.}
\label{mock-surv-prc-table}
\end{table}

\subsubsection{Utility of decision threshold}
\label{Utility of decision threshold}

Figure~\ref{fig:mock-surv-prc} examines the role of the decision threshold (the "probability" above which an image is classified as a post-merger) on the purity and completeness of the out-falling post-merger sample. We also detail selected purity-threshold combinations in Table~\ref{mock-surv-prc-table}. The sigmoid activation function in the CNN's final layer assigns each image a value between 0 and 1, roughly representing the model's certainty in its classification: values very close to one correspond to high post-merger certainty, and values near zero correspond to high non-post-merger certainty. The decision threshold is the cut used to separate post-mergers and non-post-mergers based on the "probabilities" assigned by the model.

Calibration curves and the expected calibration error (ECE) metric are often used to assess the practical utility of a network's decision boundary as a true metric of probability, and to evaluate whether re-calibration is necessary. Using the model's predictions on the reserved test data from Section~\ref{Test Set Properties}, and following the method given in \citet{DBLP:journals/corr/GuoPSW17}, we construct a calibration curve with 15 quantile bins (i.e. an equal number of samples in each bin of the model's predictions), and calculate an ECE of \textasciitilde2\%. This result is consistent with calibration characteristics for a number of post-correction models detailed in \citet{DBLP:journals/corr/GuoPSW17}, and so we are able to proceed in using the decision boundary as a metric of post-merger certainty in good faith.

Until now, our analysis has used a default threshold of 0.5. Using a a threshold of \textasciitilde0.98, we find that reasonably large post-merger samples containing hundreds of galaxies can still be recovered with intermediate purity of 50-60\%, and a purity of \textasciitilde80\% can be achieved when one solely considers galaxies that have been assigned post-merger labels with near-absolute certainty (decision threshold of 0.999). Because there are precious few post-mergers to identify in the survey, however, any significant movement of the decision threshold sacrifices the bulk of the true positive galaxies, and may introduce biases into subsequent scientific consideration of the out-falling sample.

\begin{figure}
\includegraphics[width=\columnwidth]{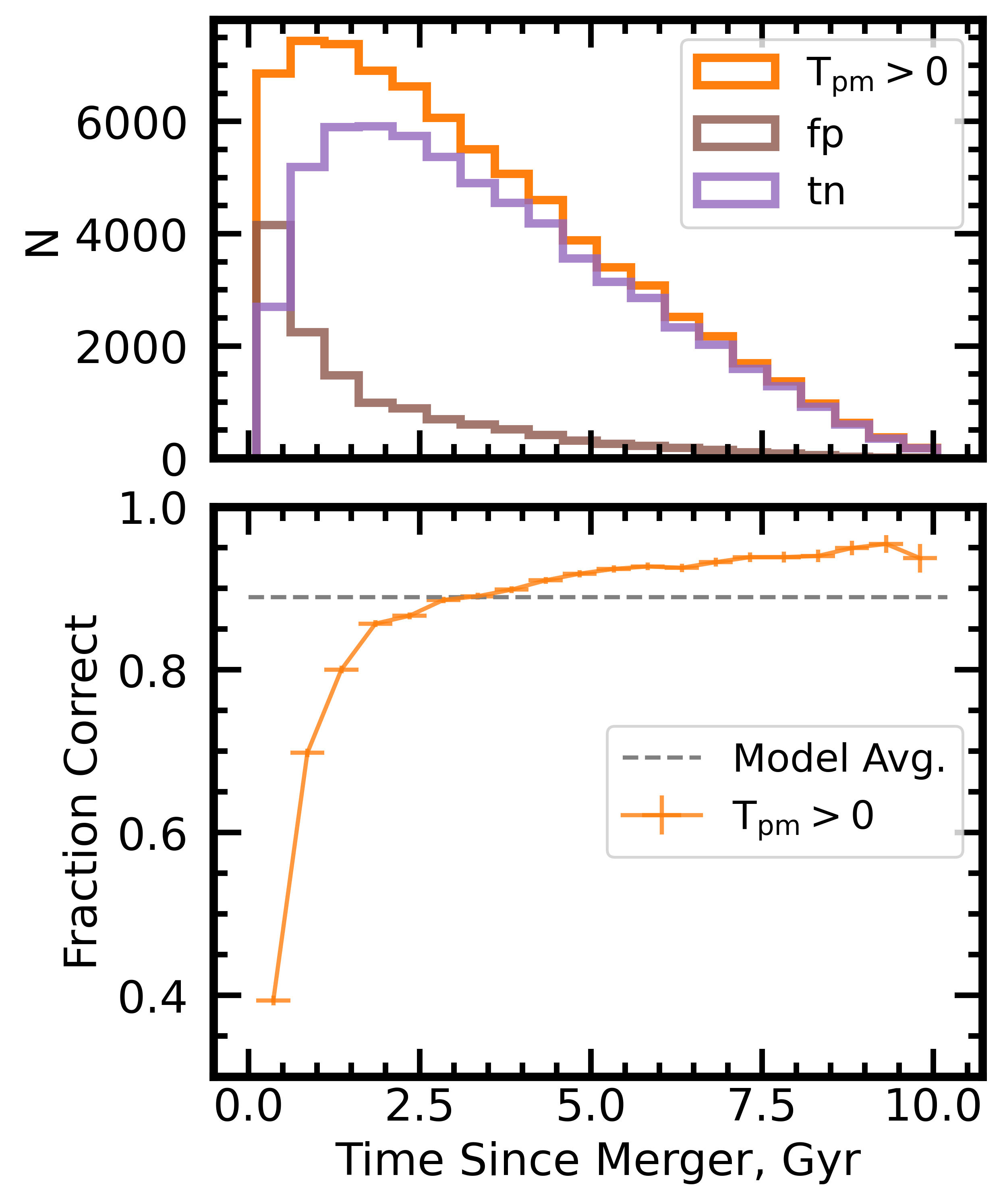}
\caption{Top panel: The number of non-post-merger objects binned by time since merger (orange histogram), and accompanying histograms for the two possible confusion matrix categories (fp, brown: non-post-mergers classified as post-mergers; tn, purple: correctly classified non-post-mergers) into which they could fall, using the default model decision threshold (0.5). Bottom panel: The fraction of correctly-classified non-post-merger objects binned in the same way.}
\label{fig:mock-surv-perf-vs-tpm}
\end{figure}

\subsubsection{Role of time-since-merger}
\label{Role of time-since-merger}

In our analysis so far, we have imposed a very strict definition of "post-merger" on the sample selection, requiring that the merger occurred in the time between the present and the previous simulation snapshot. Additionally, the non-post-merger class in the mock survey is no longer an equal-sized control group as in Section~\ref{Test Set Properties}, but rather any simulation object with a stellar mass greater than $\mathrm{10}^\mathrm{10}$ \(\textup{M}_\odot\) not explicitly counted as a post-merger. Figure~\ref{fig:mock-surv-perf-vs-tpm} examines the sensitivity of classification accuracy for the non-post-merger galaxies to the times since their most recent mergers. The sensitivity to time-since-merger for the galaxies is strong, particularly for galaxies that last merged within 2 Gyr. Galaxies that belong to the non-post-merger class by our definition and have undergone a merger in the last 0.1 Gyr are \textasciitilde60\% likely to receive a post-merger label using the default decision threshold (0.5), i.e. many galaxies are classified as post-mergers due to a real merger that has occurred in the relatively recent past. The CNN is therefore identifying bona-fide merger features that are persisting in time. As $\mathrm{T}_\mathrm{Postmerger}$ increases in non-post-merger galaxies, the chance of correct classification increases as well. Still, non-post-merger galaxies that have not experienced a merger in the last 2 Gyr stand a chance of being classified as post-mergers, and so the problem of legitimate sample impurity persists, albeit to a lesser extent.

\subsubsection{Mock survey star formation enhancement study}
\label{Star Formation Enhancement Study}

In order to assess the impact of purity and completeness on our ability to accurately quantify the physical changes incurred by a merger, we investigate the star formation rates (SFRs) of post-mergers in IllustrisTNG. The objective is to compare the true enhancement in SFRs exhibited by the full post-merger sample in TNG (e.g. as previously quantified by \citealp{2020MNRAS.493.3716H}) with the recovered SFR enhancement exhibited by the machine-predicted out-falling post-merger sample identified by the CNN.

To compute a given galaxy's SFR enhancement in TNG 100-1, we pre-select star-forming galaxies. To this end, we fit a line to the star formation main sequence (SFMS), or the correlation of star formation with stellar mass (e.g. \citealp{2014ApJS..214...15S}) in each simulation snapshot, and measure the perpendicular scatter of galaxies about the line in order to apply a snapshot-wise cut 2-$\sigma$ below it. This cut removes quiescent galaxies from both the potential post-merger and control pools, in order to facilitate analysis of star formation enhancement within a population of already star-forming galaxies, setting aside the question of merger-induced galaxy revivification. Subsequently, we again follow \citet{2020MNRAS.493.3716H} and identify post-merger galaxies within the star-forming population as in Section~\ref{Post-mergers}, and control eligible non-post-mergers as having a $\mathrm{T}_\mathrm{Postmerger}$ of $\ge2 \mathrm{Gyr}$. The resulting post-merger sample consists of 971 galaxies (i.e. 1361 post-mergers are removed from the sample due to inadequate star formation), while the control pool includes 126,577. In order to quantify the effects of the merger, we search for control galaxies using two different methods:
\begin{enumerate}
    \item matching on stellar mass and simulation lookback time only, and
    \item matching on $\mathrm{r_{1}}$ and $\mathrm{N_{2}}$ in addition to stellar mass and simulation lookback time.
\end{enumerate}
Method (i) mimics an observationally-driven approach, where accurate statistics about a galaxy's environment may be challenging to measure due to spectroscopic incompleteness (e.g. \citealp{2008ApJ...685..235P}), while method (ii) seeks to carefully account for nearby neighbours and extended environment in order to study star formation enhancement with as few biases as possible. The default control tolerance for stellar mass is 0.1 dex, and the default control tolerance for $\mathrm{r_{1}}$ and $\mathrm{N_{2}}$ are both 10\%. Controls must be drawn from the same simulation snapshot, and so there is effectively no lookback time tolerance; this represents a small deviation from Section~\ref{Controls-Environment} in order to avoid applying incongruous SFMS criteria to a given post-merger and its controls. If the default tolerances yield a control pool of more than five galaxies, their median star formation rate is subtracted from that of the post-merger in question in order to calculate an individual $\Delta$SFR. If there are five or fewer eligible controls, the tolerances are loosened as they are in Section~\ref{Controls-Environment}. Tolerances are allowed to grow a maximum of four times, in order to maintain reasonable resemblance between post-mergers and controls.

\begin{figure}
\includegraphics[width=\columnwidth]{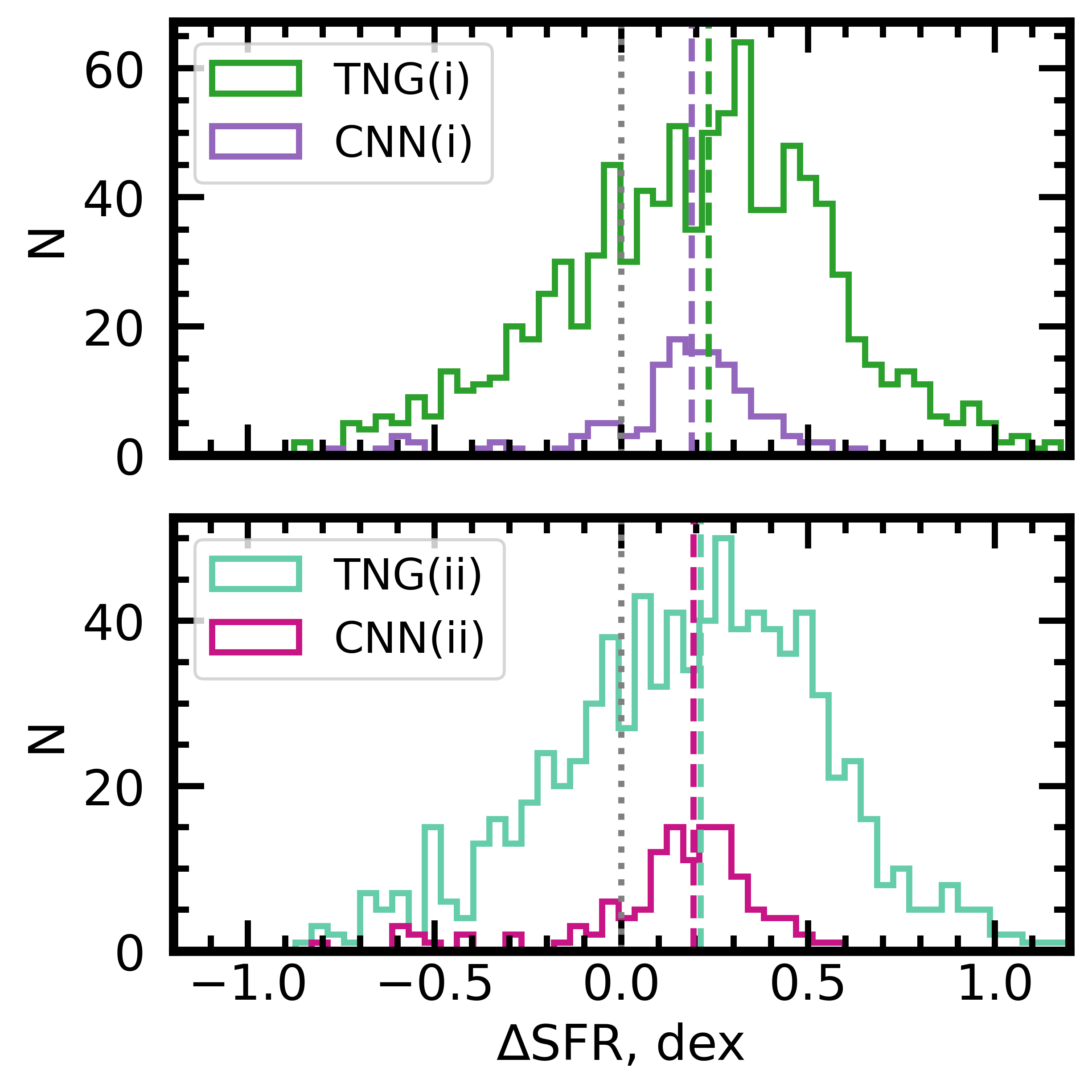}
\caption{Star formation enhancement histograms for control methods (i) and (ii), top panel and bottom panel, respectively. Post-mergers identified using Illustris-TNG metadata with controls matched in stellar mass and redshift only shown in green. Post-mergers for the turquoise curves are identified in the same way, but controls are found using Illustris-TNG environmental parameters ($\mathrm{r_{1}}$ and $\mathrm{N_{2}}$) as well. The violet and magenta curves are control-matched in the same way, but are identified by the CNN using extreme decision thresholds for both classes (0.001 for controls, and 0.999 for post-mergers).}
\label{fig:mock-surv-dsfr}
\end{figure}

Figure~\ref{fig:mock-surv-dsfr} shows that our approach yields a "true" value $\Delta$SFR of 0.23 dex for method (i), which included all post-mergers, and 0.21 dex for method (ii), which was able to find suitable controls for 857 of the post-mergers. The green (top panel) and turquoise (bottom panel) histograms in Figure~\ref{fig:mock-surv-dsfr} correspond to methods (i) and (ii), respectively.

We then repeat the calculation of $\Delta$SFR, but now for post-mergers that have been identifed by the CNN, rather than those selected based on the $\mathrm{T}_\mathrm{Postmerger}$ flag within the simulation itself. In order to impress purity upon the post-merger sample, we only consider galaxies that have been labeled as $\ge0.999$ by the model to be eligible post-mergers. Similarly, we only allow galaxies labeled as $\le0.001$ to be eligible controls. Combining the mock survey labels and the SFMS criterion detailed above results in a star forming CNN-predicted post-merger sample of 140 galaxies, and a control-eligible pool of 120,669. As before, method (i) found suitable controls for the full CNN-predicted post-merger sample, while method (ii) adequately controlled for all but fourteen galaxies. Figure~\ref{fig:mock-surv-dsfr} shows that although the sample of 140 CNN-predicted post-mergers is in fact only \textasciitilde49\% pure (reduced from the original 80\% after removing galaxies with inadequate star formation), both control-matching methods result in median $\Delta$SFR quantities that closely track the simulation's ground-truth values: 0.19 dex for both control methods (violet and magenta histograms in Figure~\ref{fig:mock-surv-dsfr}). Thus, the CNN-based approach recovers a qualitatively credible trend, and impurity in the post-merger sample and the low-number statistics associated with the extreme-threshold CNN approach only give rise to a small discrepancy of 0.02-0.04 dex. As for the simulation ground truth, the inclusion of environmental statistics in identifying controls for post-merger galaxies does not appear to have a meaningful effect for the CNN-identified samples. Therefore, even though the CNN is susceptible to the effects of unusually dense environments (see Figure~\ref{fig:fs-test-perf-vs-r1}), neglecting to control for environment does not strongly impact the results due to the relative rarity of IllustrisTNG galaxies in such environments (see Figure~\ref{fig:multi_hist}).

\begin{figure}
\includegraphics[width=\columnwidth]{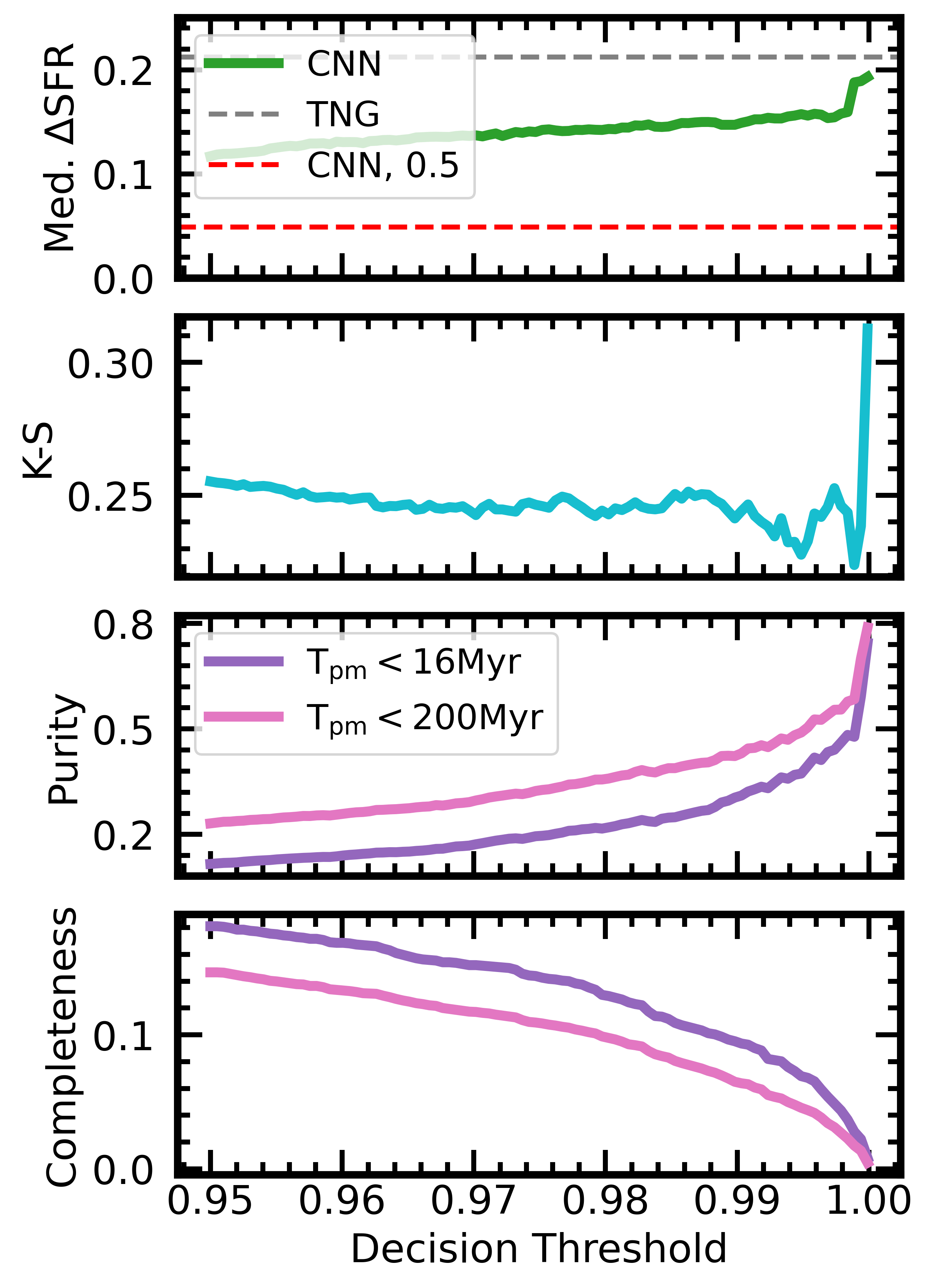}
\caption{The $\Delta$SFR in dex (top panel), K-S statistic (second panel), sample purity (third panel), and post-merger sample completeness (bottom panel) as we move the post-merger CNN decision threshold to the right, while moving the control decision threshold to the left in symmetrical fashion. Horizontal lines are included in the $\Delta$SFR panel, designating the simulation ground truth value (gray) and the CNN result using the natural decision threshold at 0.5 (red). Two purity and completeness curves are plotted, the first of which uses the strict post-merger definition (purple), while the other (pink) uses a more tolerant definition (having merged within 0.2 Gyr.)}
\label{fig:mock-surv-thresh-stats}
\end{figure}

\subsubsection{Star formation enhancement and purity}
\label{SFE and purity}

Figure~\ref{fig:mock-surv-thresh-stats} investigates the role of the decision threshold on the predicted median post-merger $\Delta$SFR in the mock survey. The  green line in the top panel of Figure~\ref{fig:mock-surv-thresh-stats} shows how the median $\Delta$SFR of the CNN-identified sample approaches the simulation ground truth (grey dashed line) as the decision threshold is changed. With the natural binary classification decision threshold in place at 0.5, the result of our mock star formation enhancement study is consistent with little to no $\Delta$SFR (red dashed line). Figure~\ref{fig:mock-surv-thresh-stats} therefore highlights that the use of a natural decision boundary in our mock survey would completely miss the statistical SFR enhancement due to high impurity. The second panel shows the Kolmogorov–Smirnov (K-S) statistic (\citealp{smirnov1948}) as a function of decision threshold, roughly representing the probability that the CNN-predicted $\Delta$SFR distribution was drawn at random from the simulation's ground truth distribution. The probability remains fairly consistent until an extreme decision threshold (i.e. the same as in Figure~\ref{fig:mock-surv-dsfr}) is applied, at which point it exceeds 0.3. Therefore, according to the K-S test, the CNN-identified $\Delta$SFR distribution is most likely to belong to the underlying IllustrisTNG distribution at the highest CNN decision threshold. The final two panels show the purity and completeness of the predicted post-merger sample as a function of threshold, with strictly-defined post-mergers ($\mathrm{T}_\mathrm{Postmerger}$ = 0) represented by the purple line, and galaxies that have merged within 0.2 Gyr shown in pink.

Without leveraging some metric of "probability", false positive galaxies, which are more numerous and less likely to have enhanced star formation than their true positive counterparts, are certain to dominate the statistics. We consider this default-threshold finding to be analogous to the recent findings of \citet{2019A&A...631A..51P}, who identify a typical merger-induced $\Delta$SFR consistent with zero. Although the training labels used by Pearson et al. are constructed using human-generated responses in Galaxy Zoo, and therefore may benefit from degrees of physical understanding that our CNN may not, we note that even the most minuscule false positive rate will give rise to a highly impure sample from Bayesian statistics, and the non-merger prior in particular (see Section~\ref{Limitations}). The lack of an enhancement in SFR in \citet{2019A&A...631A..51P} may therefore be a consequence of high impurity in the identified post-merger sample. As the model's post-merger and control thresholds are made more extreme, however, we succeed in enforcing relatively high purity in the sample. The trade-off with an extreme decision threshold is that we also rule out most post-merger galaxies and may introduce accidental biases into the post-merger sample.

\subsection{Comparison of the CNN to automated methods}
\label{Comparison to Automated Methods}

\begin{figure}
\includegraphics[width=\columnwidth]{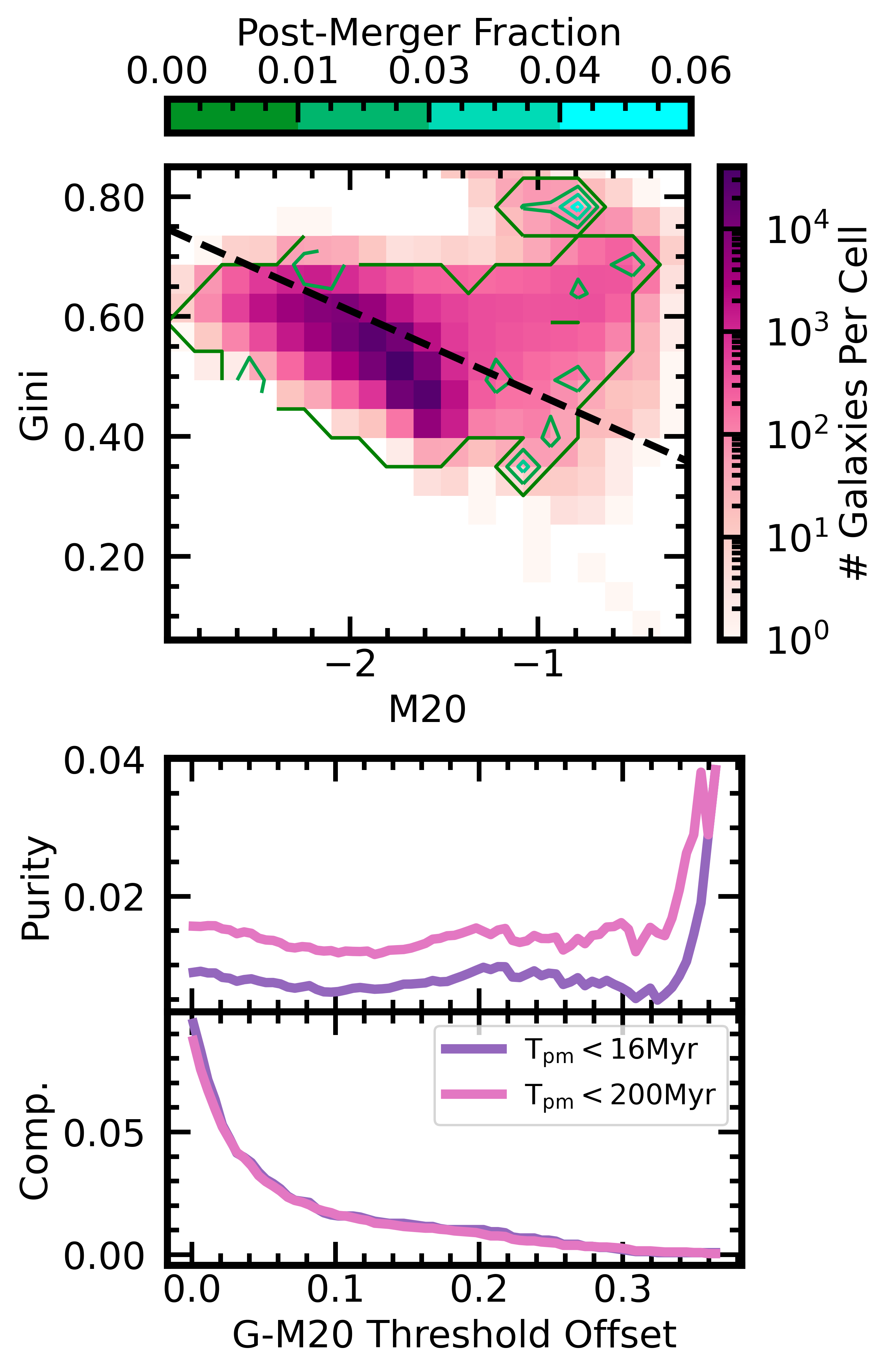}
\caption{Top panel: a 2D histogram (magenta background) of mock survey galaxies and their positions on the Gini-M20 plane, with colour in log scale. The green-blue contours plotted over the histogram show the post-merger fraction. The black dashed line shows the default threshold position, used by \citealp{2008MNRAS.391.1137L} and later \citealp{2019MNRAS.483.4140R}. Middle and bottom panels: Purity and sample completeness, respectively, as a function of the vertical shift of the Gini-M20 line shown above. The purple line applies a strict post-merger definition, and only counts galaxies that merged within the last IllustrisTNG snapshot, while the pink line uses a 0.2 Gyr cutoff.}
\label{fig:gm20-comp}
\end{figure}

\begin{figure}
\includegraphics[width=\columnwidth]{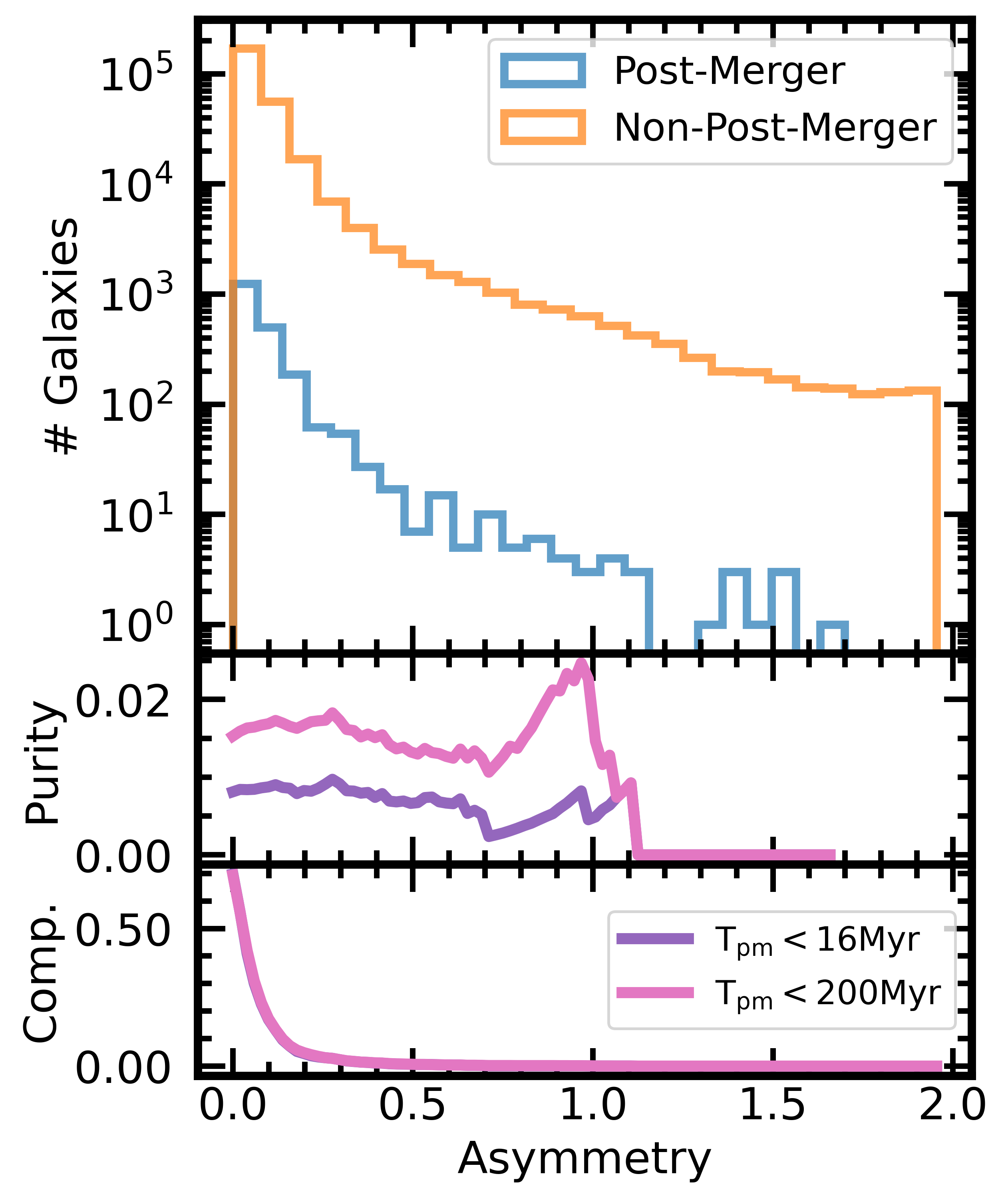}
\caption{Top panel: Log scale histograms of the post-merger and non-post-merger populations in asymmetry as measured by StatMorph. Middle and bottom panels: Purity and sample completeness, respectively, as a function of the asymmetry threshold. The purple line applies a strict post-merger definition, and only counts galaxies that merged within the last IllustrisTNG snapshot, while the pink line uses a 0.2 Gyr cutoff.}
\label{fig:A_comp}
\end{figure}

Like the CNN, several non-parametric morphological measurements are also tuneable, and have been used in post-merger identification efforts. After an image sample is automatically evaluated, optimal post-merger identification thresholds can be chosen based on the perceived purity of the out-falling sample. One popular method uses a cut applied in the Gini-M20 plane, given by \citet{2008MNRAS.391.1137L}:

\begin{equation}
Gini > -0.14\times M20 + 0.33
\label{Gini-M20}
\end{equation} where galaxies above the line are counted as mergers. A critical threshold in asymmetry has also been used, as in \citet{Conselice_2003}:

\begin{equation}
A > 0.35
\label{Asymmetry}
\end{equation} where galaxies with large $A$ values are counted as mergers, or alternatively in shape asymmetry, defined by \citet{2016MNRAS.456.3032P}:

\begin{equation}
A_s > 0.2
\label{Shape Asymmetry}
\end{equation} where 0.2 is used as the the lower bound for merger identification in SDSS. The optimal choice for CFIS may be entirely different (e.g. Wilkinson et al., in prep), and we therefore consider a range of possible boundaries in Figures~\ref{fig:gm20-comp} and ~\ref{fig:A_comp}.

In addition to their utility for rapid classification of large numbers of images, these non-parametric methods, like the CNN-based approach, can be modulated: the Gini-M20 plane cutoff can be shifted to a more severe position, and the asymmetry and shape asymmetry thresholds can be calibrated for a given data set. To compare the CNN's classification abilities to these other, more direct statistical methods, we process the \textsc{RealSim-CFIS} images generated for the mock survey (Section~\ref{Mock Survey}) using StatMorph\footnote{statmorph.readthedocs.io/en/latest/} (\citealp{2019MNRAS.483.4140R}) in order to obtain measurements for Gini and M20, as well as asymmetry and shape asymmetry. We then compare the combinations of post-merger purity and sample completeness obtained using non-parametric methods as a function of their respective thresholds to those obtained by the CNN. For all tests that make use of StatMorph, we discard galaxies for which StatMorph raised an error of any kind during analysis. This does not disproportionately impact the post-merger or non-post-merger population; around 26\% of the images belonging to each class were flagged. The sample considered hereafter therefore contains 1,734 post-merger galaxies according to the strict ($\mathrm{T}_\mathrm{Postmerger}$ = 0 Gyr) definition, and 218,536 non-post-mergers.

Figure~\ref{fig:gm20-comp} applies the Gini-M20 plane approach to the synthetic images generated for this work to reasonably good effect, given the natural over-representation of non-post-merger objects in the data. The top panel of Figure~\ref{fig:gm20-comp} shows the distribution of mock survey galaxies in the Gini-M20 plane. Green contours illustrate post-merger fraction, and the dashed line shows the cut (Eqn.~\ref{Gini-M20}) proposed by \citet{2008MNRAS.391.1137L} to distinguish the locus of post-mergers in the mock sample.  However, the relative position of the dashed line and green contours (top panel of Figure~\ref{fig:gm20-comp}) indicates that shifting this cut upwards could increase sample purity. While post-mergers are unusually abundant in the spur sequence in the upper-right of the figure, visual inspection of the spur galaxies reveals that foreground stars and other CFIS artifacts are also present in disproportionate quantities. The lower two panels of Figure~\ref{fig:gm20-comp} (for Gini-M20) are analogous to the lower two panels in Figure~\ref{fig:mock-surv-thresh-stats} (for the CNN) and show how the purity and completeness of the post-merger sample change as a function of threshold.  As for Figure~\ref{fig:mock-surv-thresh-stats}, the lower two panels in Figure~\ref{fig:gm20-comp} show the results for both post-mergers that are in their first post-coalescence snapshot (purple lines) and those that have merged in the last 0.2 Gyrs (pink lines). Shifting the Gini-M20 cut upwards leads to increased purity, but consequently the post-merger sample grows increasingly incomplete, leaving fewer galaxies for subsequent study. We therefore find that an IllustrisTNG galaxy's Gini-M20 plane position does bear on its likelihood to have undergone a recent merger, but report weaker performance in both sample purity and completeness compared to the CNN. At the default setting for each method, i.e. the CNN decision threshold at 0.5 and the Gini-M20 cutoff at the position detailed in \citealp{2008MNRAS.391.1137L}, the CNN returns a sample that is 6\% pure, while the Gini-M20 sample is not meaningfully distilled from the natural occurrence rate of 0.7\%. The sample identified using Gini-M20 is also much more incomplete, because galaxies of all types, including post-mergers, are most likely to fall below the line (towards the bottom left of the top panel in Figure~\ref{fig:gm20-comp}). After modulating both methods to their most extreme thresholds, Gini-M20 can be used to produce an enhanced purity of \textasciitilde4\%, though in a sample that is negligible in size (<1\% complete), while the CNN can produce a sample that is \textasciitilde80\% pure (an order of magnitude more effective than Gini-M20), while successfully recovering ~5\% of the sample. In short, the Gini-M20 method yields a post-merger sample that is a factor of 20 less pure than the CNN, even when pushed to its maximal performance.

Similarly, Figure~\ref{fig:A_comp} examines galaxy asymmetries in our mock survey. In the top panel, we show the distribution of asymmetry values (following the application of \citealp{Conselice_2003} within StatMorph) for post-mergers (blue) and non-post-mergers (orange). Both categories exhibit a similar overall range in asymmetry. In the lower two panels of Figure~\ref{fig:A_comp} we again show the purity and completeness of the out-falling post-merger sample as a function of asymmetry threshold. Unlike Gini-M20, the purple line in the middle panel of Figure~\ref{fig:A_comp} demonstrates that a simple asymmetry threshold cannot be used to enhance the purity of the strictly-defined post-merger sample in the mock survey. As the sample completeness decreases, purity hovers around its natural value of \textasciitilde0.7\% before dropping near $A=1$ when all post-mergers have been ruled out by the threshold. A temporary enhancement in the purity of the softer-defined post-merger sample is achieved, however, near $A=1$, though it never exceeds 3\%. In addition to asymmetry (Figure~\ref{fig:A_comp}) we investigated shape asymmetry, which is also provided by StatMorph. As with asymmetry, shape asymmetry also yields low sample purity (at most, \textasciitilde0.8\%) which is not improved by varying the threshold. Asymmetry (and shape asymmetry) therefore yield a predicted post-merger sample that is even more impure than that of Gini-M20. Visual inspection of the sample suggests that foreground stars or other survey artifacts are frequently responsible for high asymmetries. Since these phenomena do not preferentially affect post-mergers or non-post-mergers, it is reasonable that post-merger purity would not necessarily be enhanced by asymmetry thresholds in excess of $A=1$.

For the mock survey, therefore, a CNN trained with synthetic image data shows much stronger performance than any of the three popular merger identification methods detailed, with a post-merger sample an order of magnitude more pure. While the CNN benefits from calibration and training on images similar to those studied in the mock survey, it is likely that its success on this task also speaks to a diversity in the morphological characteristics of simulated post-mergers. A trainable, highly-dimensional statistical tool like the CNN may be better prepared to characterize the features relevant to post-merger status, and recognize them when applied to new images.

\subsection{Comparison of the CNN with visual classifications}
\label{Visual Classification}

Having demonstrated the CNN's ability to out-perform non-parametric methods of post-merger identification in the context of our mock survey, we next turn to visual classification. CNNs identify images in much the same way that a human being does: by learning to recognize features and feature combinations that are relevant to the image's classification. Additionally, trained CNN models require only a fraction of a second to assign a label to an image, while people may require up to several minutes to rigorously classify a single image. Human visual classifications are not without merit, however; human classifications benefit from inherent physical intuition, and in some cases, degrees of relevant training in image classification, all of which lend priors that affect the method by which an individual approaches a classification problem.

In order to compare the performance of the machine to that of a group of people, we organize a visual post-merger classification exercise using a 200-image subset of the data. Given the size of the mock survey, it was not practical to visually inspect the full data set, and so we have selected a small subset for the purpose of a straightforward yet lightweight comparison. The sample is constructed in pseudo-random fashion so that the authors can also participate in good faith, allowing for between 30\% and 70\% post-mergers by construction. In practice, the visual classification sample contains 112 post-mergers and 88 control images arranged in random order. The participant group (which includes co-authors RWB, CB, SLE, SW, and the volunteers recognized in the acknowledgments) are provided with no information about the post-merger occurrence rate in the image sample, and are asked to assign either a post-merger or non-post-merger label to each image. The classifications are marked in a standardized text file containing a list of galaxy identification numbers, each corresponding to the filename of a single galaxy image. Images are provided in Flexible Image Transport System (FITS) format so that popular astronomy visualization tools can be applied. Because we intend to compare the visual classification results to those of the machine, the participants are instructed to strive for total correctness in classification, as the CNN does by design, and not necessarily post-merger sample purity.

\subsubsection{Classifier statistics}
\label{Classifier Statistics}

\begin{figure}
\includegraphics[width=\columnwidth]{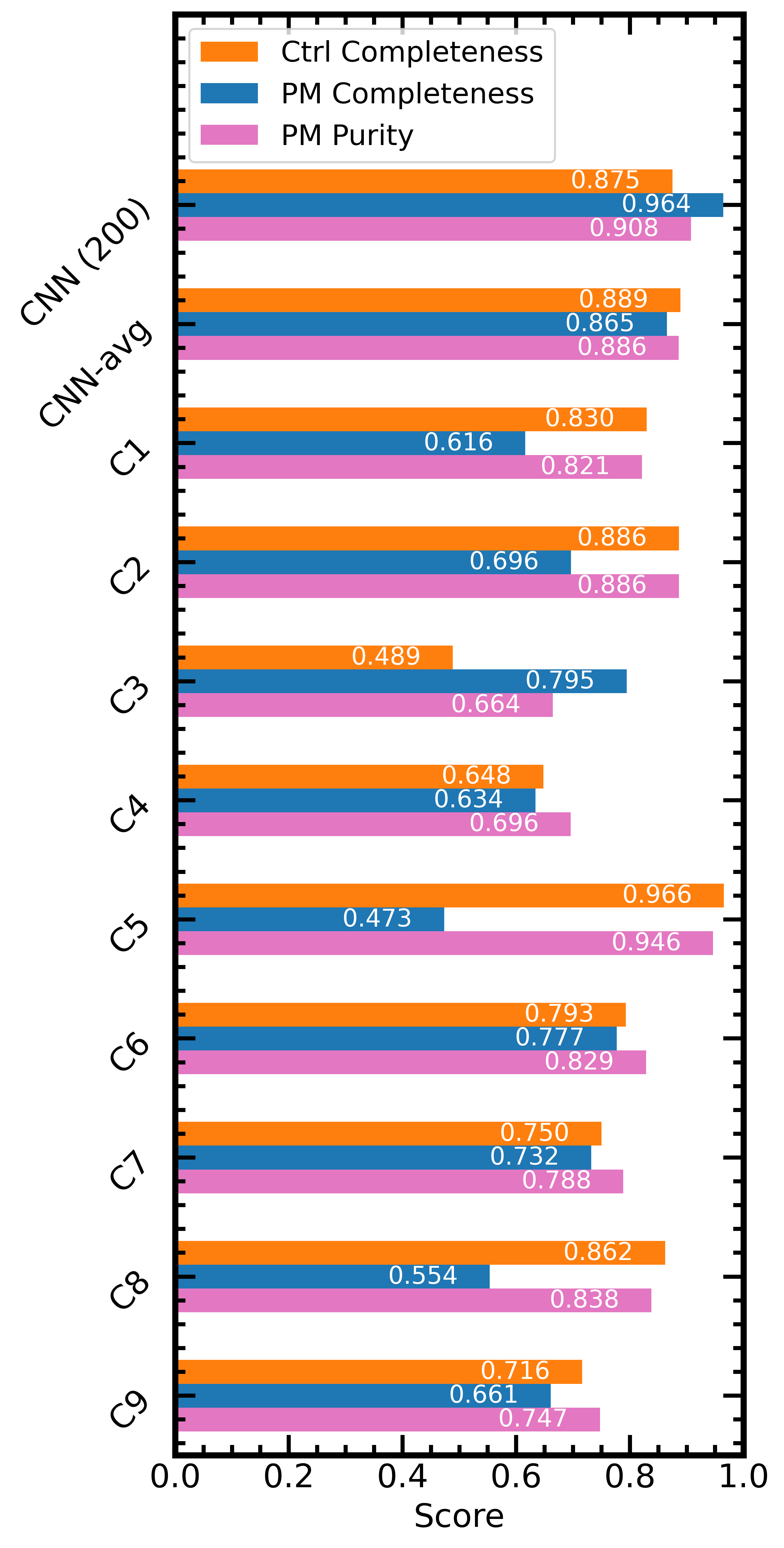}
\caption{The post-merger sample completeness, control completeness, and post-merger purity statistics for various contributors.}
\label{fig:vis-class-lb}
\end{figure}

Figure~\ref{fig:vis-class-lb} shows the fraction of control galaxies and post-mergers that were correctly identified by the CNN at the default decision threshold, and by the human classifiers. The fractional purity of each classifier's final post-merger sample is also shown. CNN (200) refers to the specific performance of the machine on the 200 selected images, while CNN-avg refers to the CNN's average performance, as detailed in Figure~\ref{fig:fs-test-cmx}. Classifiers 1, 2, and 3 are astronomy faculty members, classifiers 4 and 6 are pursuing postdoctoral studies in astronomy, 5 and 7 are graduate students in astronomy, classifier 8 is an undergraduate student in astronomy, and classifier 9 is a mechanical engineer who was given a short lecture on the properties of post-merger galaxies and their manifestations in images, using selected images from the training set that were not included in the classification task as examples.

In general, the total experience with galaxy image classification (i.e. the total number of images classified in an individual's career) seems to have a bearing on each person's performance, but other factors may also be at play. For instance, participants experienced with synthetic images (e.g. C6, C7) are more likely to have a CNN-like balance of statistics on the test data, while those more familiar with survey observations (e.g. C1, C5) are more conservative in their assignment of post-merger labels, and exchange post-merger sample completeness for purity. Generally, the CNN recovers a higher fraction (96\%) of the post-mergers than does any human participant. Even when instructed to focus on total accuracy, however, several human classifiers excel in post-merger sample purity, in some cases approaching or even surpassing (e.g. C5) that of the CNN for this image subset.

\subsubsection{A hybrid approach}
\label{Hybrid Approach}

The visual classification experiment can also be leveraged as a test of the combined classification power of a CNN algorithm and one or more individuals. When applied to a large survey, here approximated by the mock survey detailed in Section~\ref{Mock Survey}, the excellent ability of a trained CNN to identify individual post-mergers allows it to generate a predicted post-merger sample that is highly complete and more pure than the original survey. While doing so, it simultaneously de-prioritizes the visual inspection of a large majority of previously-unclassified images. Our visual classification experiment suggests that human classifiers are generally not as good at identifying a complete sample, but benefit from a physical understanding, and can excel in the identification of pure post-merger samples. All participants were able to enhance the purity of the post-merger sample when considering only galaxies assigned post-merger labels by the machine, in most cases by a few points, and up to 100\% in the case of C2. In addition, since the participants were instructed to approach the problem with an interest in total performance rather than purity, an adjustment in the purpose and parameters of future classification tasks would further enhance this effect when applied to survey data.

\subsection{The Bayesian disadvantage}
\label{Limitations}

Our IllustrisTNG mock survey in Section~\ref{Mock Survey} serves as a useful test of the recovery and study of post-merger galaxies in the Universe, in particular with respect to CFIS, around which the parameters of our synthetic observations are molded. Provided that the physics implemented in IllustrisTNG produce post-merger and non-post-merger galaxies that are visually characteristic of low-redshift survey galaxies in the Universe, the scarcity of post-mergers in the real Universe still poses a challenge to the purity of the outfalling sample, which can be quantified by Bayes rule (\citealp{1763RSPT...53..370B}):
\begin{equation}
\begin{aligned}
 \Pr(A|B)=\frac{P(B|A)P(A)}{P(B|A)P(A)+P(B|\neg A)P(\neg A)}
 \label{Bayes}
\end{aligned}
\end{equation}
where $A$ can be thought of as an intrinsic state, and $B$, for our purposes, is a prediction or indication. $Pr(A|B)$, therefore, is the likelihood of some intrinsic state $A$ given a prediction $B$, $P(B|A)$ is the probability of the prediction $B$ given the intrinsic state $A$, $P(A)$ is the intrinsic probability of the state $A$, $P(B|\neg A)$ is the probability of the prediction $B$ when the intrinsic state is something other than $A$, and $P(\neg A)$ is the intrinsic probability that the state is something other than $A$. Within the specific context of post-merger identification, we can quantify the probability of post-merger status ($A$) given a post-merger label ($B$) of any origin, whether from a human classifier or a machine. In this work, we achieve a post-merger identification rate of \textasciitilde89\%, and so $P(B|A)$, the fraction of correctly-identified post-mergers given by $P(A|B) P(B) / P(A)$, is 0.89. $P(B|\neg A)$, the fraction of incorrectly-classified non-post-mergers given by $P(\neg A | B) P(B) / (1-P(A))$, is ~11\%, or 0.11. Our greatest disadvantage lies in the values of $P(A)$ and $P(\neg A)$, the natural occurrence rates of post-mergers and non-post mergers in the Universe (0.007 and 0.993 in TNG100-1). In total, Bayesian statistics predict that an object labeled as a post-merger is only \textasciitilde5\% likely to belong to the post-merger class. Increasing the decision threshold serves to inflate the value of $P(B|A)$, but greatly reduces post-merger completeness. Further, as demonstrated in Section~\ref{Star Formation Enhancement Study}, additional cuts applied during any study of the out-falling sample of post-mergers may unwittingly reduce the sample purity, even after a strict decision threshold has been applied.

This challenge does not only apply to automated and machine learning model classifications, however. Section~\ref{Visual Classification} reviews the performance of the machine and several individuals on a small subset of the image sample. Even though two individuals were able to either match or surpass the CNN's post-merger purity metric, any individual classifier with a post-merger sample that is less than 100\% pure will return results that are significantly contaminated as a result of the minuscule value of $P(A)$.

\section{Summary}
\label{Summary}

Large samples of pre-coalescence galaxy pairs have facilitated detailed and statistically robust study of the pair phase of galaxy mergers. However, since post-coalescence galaxies are more difficult to identify via automated methods, large samples of post-merger galaxies are missing from the literature. The details of this transformative period in galaxy evolution therefore remain relatively unstudied.

The Canada-France Imaging Survey (CFIS) will contain relatively deep, excellent quality r-band imaging over \textasciitilde5000 square degrees of the sky, making it an excellent data set for the identification of recently merged galaxies. In order to avoid performing labourious visual classifications of a prohibitively large population of CFIS galaxies, we seek to develop a software tool for post-merger sample distillation in order to augment the efficiency of post-merger identification efforts.

In this paper we construct synthetic CFIS observations of simulated IllustrisTNG galaxies in order to study the merits of automated post-merger identification with a convolutional neural network (CNN). Our main findings are as follows:

\begin{itemize}

\item We train a convolutional neural network on a large sample of synthetic galaxy observations, produced by applying CFIS-motivated observational realism to the stellar mass maps obtained from 0 < z < 1 galaxies in the 100-1 run of IllustrisTNG. The final data set contains 37,312 images each of post-merger galaxies and control galaxies. We evaluate the model on a sample of post-merger and control galaxies never seen in training. It recovers 88.9\% of the control galaxies and 86.5\% of the post-mergers (Figure~\ref{fig:fs-test-cmx}), with an AUC score of 0.95 (Figure~\ref{fig:fs-test-roc}).

\item The model's performance is not significantly impacted by most relevant metadata quantities, including stellar mass and mock observation redshift (Figures~\ref{fig:fs-test-perf-vs-mstar} and \ref{fig:fs-test-perf-vs-z}). The model is most prone to misclassification when a galaxy has one or more nearby neighbour(s) within a few tens of kpc (Figure~\ref{fig:fs-test-perf-vs-r1}). In these cases, the visual characteristics of the morphological disturbances induced by the partner galaxy / galaxies are highly degenerate with those associated with a recent merger, and may be indistinguishable to the network.

\item We apply the trained model to a mock survey, containing one image of every 0 < z < 1 IllustrisTNG galaxy with stellar mass >$\mathrm{10}^\mathrm{10}$ \(\textup{M}_\odot\). The post-mergers as defined in training, which are re-observed from a new camera angle and with new mock observation parameters, constitute less than 1\% of objects meeting this mass criterion. The model correctly labels \textasciitilde89\% of both the post-mergers and the non-post-mergers in the survey (Figure~\ref{fig:mock-surv-cmx}). However, from Bayes theorem, the scarcity of post-mergers relative to non-mergers means that the out-falling post-merger sample is only 6\% pure when the default CNN decision threshold is used (Figure~\ref{fig:mock-surv-prc}). While legitimate sample impurity exists, we note that many of the contaminating galaxies have undergone a recent merger within 0.2 Gyr (Figure~\ref{fig:mock-surv-perf-vs-tpm}).

\item We use the CNN's labels and decision threshold to identify post-merger and control samples for a proof-of-concept study of star formation enhancement in the style of \citet{2020MNRAS.493.3716H}. In particular, we investigate how changes in the decision threshold, which affect purity and completeness of the post-merger sample, affect the SFR statistics. An extreme decision threshold scheme of $\ge0.999$ for post-mergers and $\le0.001$ for controls gives rise to 49\% purity in the post-merger sample, and returns a prediction for the median $\Delta$SFR in excellent agreement (within 0.02 dex) of the simulation's ground truth for the most conservative control-matching scheme (Figures~\ref{fig:mock-surv-dsfr} and~\ref{fig:mock-surv-thresh-stats}).

\item We compare the CNN's performance on the survey to extant non-parametric statistical methods, and find a linear cutoff in the Gini-M20 plane to be modestly effective in enhancing post-merger sample purity (up to \textasciitilde4\%, with <1\% completeness) in the mock data. (Figure~\ref{fig:gm20-comp}). However, the CNN out-performs this approach at its default decision threshold, and can achieve an incredibly pure sample (\textasciitilde80\%) with better completeness (\textasciitilde5\%) at its most extreme setting. Using an asymmetry threshold failed to distill the sample any higher than 0.7\%, the natural occurrence rate of strictly-defined post-mergers in IllustrisTNG100-1, but did slightly enhance the density of galaxies with a $\mathrm{T}_\mathrm{Postmerger}$ of <0.2 Gyr to 2\% from a natural density of 1\% (Figure~\ref{fig:A_comp}). Shape asymmetry was ineffective in enhancing the purity of post-mergers in the mock survey. We also consider human visual classifications, and identify a tradeoff: visually-identified samples can be more pure (up to 95\% in the case of Classifier 5) than those returned by the CNN, but are typically more incomplete (47\% for the same classifier, versus 96\% for the CNN on the same data) (Figure~\ref{fig:vis-class-lb}).

\end{itemize}

In Section~\ref{Hybrid Approach}, we argue that selection biases and sample impurity will be inherent in any effort, automated or manual, to identify post-mergers in a survey with a natural post-merger incidence rate, and that CNN-based post-merger identification is best utilized as a first round of distillation, free from preconceptions other than those inherent to the training data. Once a set of CFIS galaxies are processed in this way, a trained person can further improve the quality of the post-merger set through careful inspection.

The image classification techniques developed in this work have been thus far trained and tested on like-generated synthetic observations, e.g. in Section~\ref{Mock Survey}. By training a CNN on a population of simulated galaxies processed with CFIS realism, however, we have simultaneously prepared it for application to CFIS, to the extent that IllustrisTNG galaxies are morphologically representative of those found in the Universe. Thus, in future work we will apply our model to CFIS galaxies as part of a hybrid approach comparable to that detailed in Section~\ref{Hybrid Approach} in an effort to identify and study a post-merger sample of groundbreaking volume and quality.

\section*{Acknowledgements}
\label{Acknowledgements}

The work detailed above was conducted at the University of Victoria in Victoria, British Columbia, as well as in the Township of Esquimalt in Greater Victoria. We acknowledge with respect the Lekwungen peoples on whose unceded traditional territory the university stands, and the Songhees, Esquimalt and WSÁNEĆ peoples whose historical relationships with the land continue to this day.

CFIS is conducted at the Canada-France-Hawaii Telescope on Maunakea in Hawaii. We also recognize and acknowledge with respect the cultural importance of the summit of Maunakea to a broad cross section of the Native Hawaiian community.

We thank volunteers (in alphabetical order) Justin Hufnagel, David Patton, Salvatore Quai, Mallory Thorp, and Joanna Woo for their contributions to the visual classification experiment.

This work is based on data obtained as part of the Canada-France Imaging Survey, a CFHT large program of the National Research Council of Canada and the French Centre National de la Recherche Scientifique, and on observations obtained with MegaPrime/MegaCam, a joint project of CFHT and CEA Saclay, at the Canada-France-Hawaii Telescope (CFHT) which is operated by the National Research Council (NRC) of Canada, the Institut National des Science de l’Univers (INSU) of the Centre National de la Recherche Scientifique (CNRS) of France, and the University of Hawaii. This research used the facilities of the Canadian Astronomy Data Centre operated by the National Research Council of Canada with the support of the Canadian Space Agency.

Data from the IllustrisTNG simulations are integral to this work. We thank the Illustris Collaboration for making these data available to the public.

Funding for the SDSS and SDSS-II has been provided by the Alfred P. Sloan Foundation, the Participating Institutions, the National Science Foundation, the U.S. Department of Energy, the National Aeronautics and Space Administration, the Japanese Monbukagakusho, the Max Planck Society, and the Higher Education Funding Council for England. The SDSS Web Site is http://www.sdss.org/. The SDSS is managed by the Astrophysical Research Consortium for the Participating Institutions. The Participating Institutions are the American Museum of Natural History, Astrophysical Institute Potsdam, University of Basel, University of Cambridge, Case Western Reserve University, University of Chicago, Drexel University, Fermilab, the Institute for Advanced Study, the Japan Participation Group, Johns Hopkins University, the Joint Institute for Nuclear Astrophysics, the Kavli Institute for Particle Astrophysics and Cosmology, the Korean Scientist Group, the Chinese Academy of Sciences (LAMOST), Los Alamos National Laboratory, the Max-Planck-Institute for Astronomy (MPIA), the Max-Planck-Institute for Astrophysics (MPA), New Mexico State University, Ohio State University, University of Pittsburgh, University of Portsmouth, Princeton University, the United States Naval Observatory, and the University of Washington.

For their insight and comments on this work, we acknowledge Pierre-Alain Duc (Directeur Observatoire Astronomique de Strasbourg), and Sébastien Fabbro (Research Council Officer for UVic Physics and Astronomy and NRC Herzberg).

CB gratefully acknowledges support from the Natural Sciences and Engineering Research Council of Canada (NSERC).

MHH acknowledges support from the William and Caroline Herschel Postdoctoral Fellowship Fund, and the receipt of a Vanier Canada Graduate Scholarship.

This research was enabled, in part, by the computing resources provided by Compute Canada.

\section*{Data Availability}
\label{Data Availability}

Simulation data from TNG100-1 used in the generation of training images for this work are openly available on the IllustrisTNG website, at tng-project.org/data. Template versions of \textsc{RealSim} and \textsc{RealSim-CFIS}, developed by CB with modifications by RWB are publicly available via GitHub at github.com/cbottrell/RealSim and github.com/cbottrell/RealSim-CFIS. Specific image training data used to develop the findings of this study are available by request from RWB.



\bibliographystyle{mnras}
\bibliography{cnn-ill-pm} 



\bsp	
\label{lastpage}
\end{document}